\documentclass[lettersize,journal]{IEEEtran}
\usepackage{amsmath,amsfonts,amsthm}
\usepackage{algorithmic}
\usepackage{algorithm}
\usepackage{array}
\usepackage{booktabs}
\usepackage{lipsum} 
\usepackage{caption} 
\usepackage{subcaption}
\usepackage[table,xcdraw]{xcolor}
\usepackage{graphicx}
\usepackage{multirow}
\usepackage{textcomp}
\usepackage{url}
\usepackage{verbatim}
\usepackage{cite}
\usepackage[hidelinks]{hyperref} 
\usepackage{cleveref}
\usepackage{xcolor}
\definecolor{framefill}{RGB}{230, 230, 242}
\definecolor{frameline}{RGB}{60,60,60}
\newcommand{\bboxed}[1]{%
    \noindent
    \vspace{1mm}\\%
    \fcolorbox{frameline}{framefill}{%
         \begin{minipage}[]{0.98\columnwidth}
         \setlength{\parskip}{3ex}
         #1
        \end{minipage}%
    }%
    \vspace{1mm}\\%
}
\theoremstyle{definition}
\newtheorem*{definition}{Definition}
\begin{document}
\title{Leveraging Neural Graph Compilers in Machine Learning Research for Edge-Cloud Systems}
%
%
%
\author{
\thanks{Alireza Furutanpey is with Coovally Barcelona, Spain, and the Distributed Systems Group, TU Wien, Vienna.}
\thanks{Philipp Raith is with Coovally Barcelona, Spain, and the Distributed Systems Group, TU Wien, Vienna.}
\thanks{Carmen Walser is with the Distributed Systems Group, TU Wien, Vienna.}
\thanks{Pantelis A. Frangoudis is with the Distributed Systems Group, TU Wien, Vienna.}
\thanks{Schahram Dustdar is with ICREA Barcelona, Spain, and the Distributed Systems Group, TU Wien, Vienna.}
Alireza Furutanpey$^*$,
\thanks{\textsuperscript{*}Corresponding author: a.furutanpey@coovally.ai}
Carmen Walser, Philipp Raith, Pantelis A. Frangoudis, Schahram Dustdar
}
%
%
\maketitle
\begin{abstract}
This work presents a comprehensive evaluation of neural network graph compilers across heterogeneous hardware platforms, addressing the critical gap between theoretical optimization techniques and practical deployment scenarios. We demonstrate how vendor-specific optimizations can invalidate relative performance comparisons between architectural archetypes, with performance advantages sometimes completely reversing after compilation. Our systematic analysis reveals that graph compilers exhibit performance patterns highly dependent on both neural architecture and batch sizes.
Through fine-grained block-level experimentation, we establish that vendor-specific compilers can leverage repeated patterns in simple architectures, yielding disproportionate throughput gains as model depth increases. We introduce metrics 
to quantify a compiler's ability to mitigate performance friction as batch size increases. 
The resulting methodology turns graph-compiler benchmarking into design feedback for ML systems, allowing researchers to validate whether architectural, batching, and resource-utilization conclusions remain valid after compilation on target hardware.
\end{abstract}
\begin{IEEEkeywords}
Deep Learning, Neural Networks, Compilers, Graph Optimization, Benchmarking
\end{IEEEkeywords}
\section{Introduction} \label{sec:intro}
\begin{figure}[t]
    \centering
    \includegraphics[width=1\linewidth]{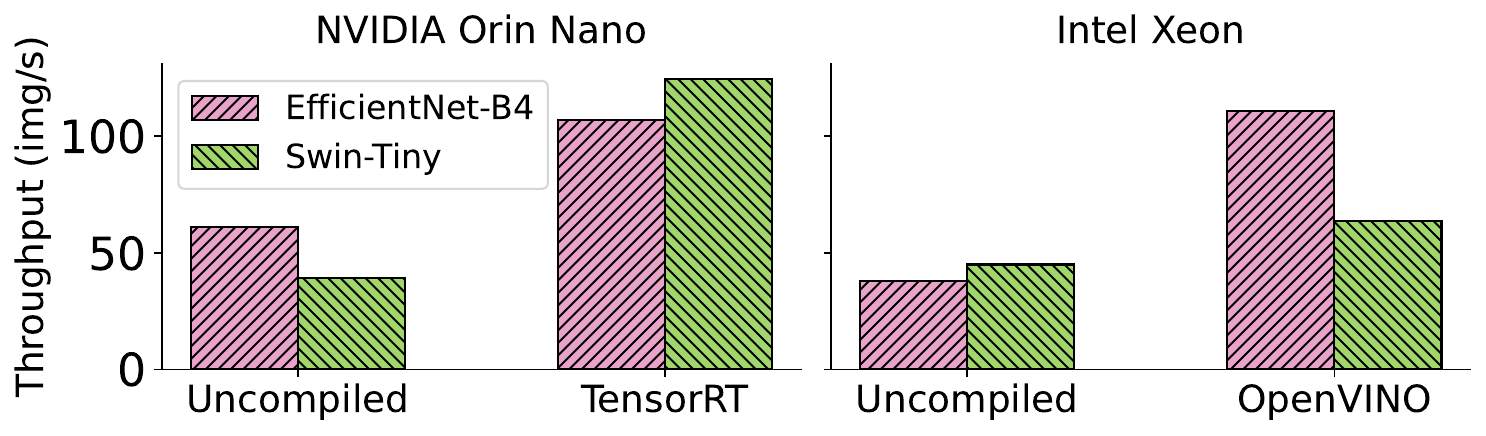}
    \caption{On the Orin, the convolutional-based EfficientNet has higher throughput than the transformer-based Swin. Applying TensorRT significantly improves throughput for both models. However, the EfficientNet is now slower than the Swin.
    On the Xeon with OpenVINO, we observe the exact inverse behavior. 
    }
    \label{plot:reversal_pilot}
\end{figure}

\IEEEPARstart{T}{he} pervasiveness of Artificial Neural Networks (ANNs) in modern computing systems has generated significant demand for methods to improve the efficiency of available hardware. 
As computational complexity increases and deployment scenarios diversify, optimizing ANN execution becomes indispensable for practical applications across various computational platforms. 
A common problem when extending systems research into real-world environments is determining whether reported performance improvements,  regarding resource usage or throughput, from the latest advancements will generalize to the target platform. This problem stems not from a lack of rigor by researchers but from the inherent heterogeneity in hardware ~\cite{reuther2022ai} and environmental factors~\cite{electronics9122106}. 
Among the most promising optimization approaches to accelerate inference are graph compilers, which optimize the computational graphs of ANNs to enhance scheduling, improve data flow, and exploit dedicated hardware modules.  
Yet, fully leveraging graph compilers presents distinct challenges, further disconnecting academic research from real-world applicability, despite directly addressing practical problems. We find that when designing novel machine learning algorithms for Edge(-Cloud) Systems~\cite{confluence}, it is crucial to understand how graph compilers can invalidate \textit{relative} performance differences between architectural archetypes. As new architectures emerge, their realized performance depends on compiler support. Comparative claims that ignore compilation risk are misleading on deployed hardware due to diverging optimization stacks, irrespective of whether the computational capacities between test beds and real-world environments are comparable. 
In particular, graph compilers and other vendor-specific optimizations can completely alter the relative performance across competing architectures. 
\Cref{plot:reversal_pilot} not only precisely exemplifies this behavior, but also demonstrates that the exact inverse holds for a different device-compiler pair 
on two models with comparable parameter counts at batch size 8
(see \Cref{sec:eval} for details on experiment configurations). This insight was a key motivation in our previous works \cite{frankensplit, fool, svbirobustness}, where we deliberately opted for simplified encoder architectures with widely supported operations to ensure that reported results would generalize across vendors. While these and similar research contributions are valuable, their practical application requires further consideration, often creating a needlessly high barrier for practitioners who must navigate complex optimization landscapes.
%
To narrow the gap between research contributions and their applications in real systems, we introduce an automated tool that integrates with existing profilers commonly used in edge or cloud frameworks (e.g., for model selection~\cite{inferline}). The tool streamlines graph compiler benchmarking over heterogeneous compute infrastructure to facilitate iterative empirical analysis.

We do not propose a new neural architecture or learning algorithm. The aim is to motivate and introduce a methodology that instruments the ML workflow with compiler effects, supported by quantitative metrics, so that researchers can determine whether architectural choices, block-composition decisions, batching assumptions, and resource-utilization claims remain valid after compilation across heterogeneous devices. 
The methodology describes using such tools to design, implement, and deploy experiments in research on Edge or Edge-Cloud systems that rely on ANNs for function approximation. It facilitates architectural comparison, compiler-device selection, and deployment-oriented evaluation as connected decisions rather than as isolated benchmarking tasks.
Additionally, we perform a comprehensive empirical analysis across varying architectural families on a heterogeneous physical testbed, demonstrating how vendors prioritize optimizing different layer compositions using both vendor-agnostic (e.g., Apache TVM) and vendor-specific graph compilers (e.g., TensorRT, OpenVINO). 

While graph compiler optimizations can also be combined with model compression or acceleration methods, such as knowledge distillation~\cite{gou2021knowledge}, quantization~\cite{li2024contemporary}, and pruning~\cite{cheng2024survey}, their advantage lies in improving execution efficiency without changing the model’s prediction behavior.
Accordingly, this study examines the isolated effects of compilers without additional confounding variables. 
In summary, this work's main contributions are:
\begin{itemize}
\item Evaluating graph compilers across heterogeneous hardware platforms, demonstrating the advantages of vendor-specific optimizations.
\item Analyzing batch parallelization efficiency across architectures and compilers, revealing optimization opportunities in resource-constrained environments.
\item Demonstrating through block-level experimentation that vendor-specific compilers leverage repeated patterns in simpler compositions for disproportionate throughput gains with increased depth.
\end{itemize}

We are aware that there is no shortage of work examining the NN graph compiler landscape (\Cref{sec:relwork,subsec:background}). 
However, to the best of our knowledge, this work is the first to connect graph-compiler benchmarking to the ML research workflow itself. Rather than reporting compiler speed-ups in isolation, we analyze how compilation changes architectural comparisons, depth and width trade-offs, batch-scaling behavior, and resource-utilization bottlenecks across heterogeneous device-compiler pairs. \Cref{sec:framework} formalizes this methodological use, and \Cref{sec:eval} provides the empirical evidence and diagnostic metrics, such as batch-scaling efficiency, needed to expose these effects.
\section{Related Work} \label{sec:relwork}
%
Shuvo et al.~\cite{aiaccelerationreview} provide an excellent review on techniques for utilizing AI accelerators, but it is focused on lower-level tricks for a particular class of hardware. 
%
Zhou \& Yang~\cite{tensorrteval} benchmark TensorRT, but only on convolutional architectures.
Li et al.~\cite{dlcompilersurvey} provide a broad overview of existing compilers, but only include rudimentary evaluation on behavior in practice. 
Like our work, Xing et al.\cite{comparisononhardware} examine graph compilers on different hardware (CPUs, GPUs), but the evaluation only considers individual operations and convolutional-based architectures, without addressing important factors, such as batch size, depth, width, etc.
Conversely, this work evaluates graph compilers on various networks from varying architectural families and leverages the broad results to draw generalizable insights. 
The work by Jajal et al. \cite{interop} shares similarities in examining computational graph optimization of varying architectural styles and vendors, but the focus is on interoperability, and specifically the issues that may be encountered when converting models to ONNX. 
The work in \cite{dltesting} also examines computational graph optimization, but more generally focuses on uncovering bugs in the development cycle of systems that train and deploy deep neural networks. 
The work in \cite{codesign} shares similarity in advocating for a design strategy that is mindful of the underlying hardware acceleration. Still, it is an entirely qualitative assessment without empirical analysis. 
%
Lastly, Zhang et al.~\cite{mobilelib, mobilelib2} introduce libraries for benchmarking and provide comprehensive results, but include only mobile platforms and do not examine graph compilers.

In summary, prior work establishes that graph compilers and hardware-aware optimization are important, but it does not characterize how compiler effects propagate into ML research decisions. Existing studies either focus on a single compiler or architectural family, evaluate isolated operators, emphasize interoperability or bug finding, or provide benchmarking libraries without examining compiler-induced changes in architectural conclusions. The novel insight of this work is that compiler effects are not merely absolute speed-up factors applied after model design. They can change relative architectural rankings, alter the apparent benefit of increasing depth or width, modify batch-scaling behavior, and shift resource-utilization bottlenecks across heterogeneous device-compiler pairs. These effects directly influence the selection of model configuration in edge-cloud research workflows. Therefore, our framework and metrics turn graph-compiler benchmarking from a final performance measurement into a diagnostic tool for validating architecture, batching, and deployment assumptions.
\section{Background}\label{subsec:background}
Graph compilers analyze and optimize computational graphs that represent the operations of a neural network as nodes and the data dependencies as edges.
They provide abstraction from lower-level implementation details by converting models from high-level frameworks (such as PyTorch and TensorFlow) into hardware-agnostic intermediate representations. Moreover, they may apply transformations that improve execution speed and memory efficiency across AI accelerators, and hardware-specific code generation for low-level kernels tailored to target architectures (e.g., CUDA for NVIDIA GPUs, OpenCL for FPGAs).
\subsection{High-Level Network Architecture Organization}
\begin{figure}[htb]
    \centering
    \includegraphics[width=1\linewidth]{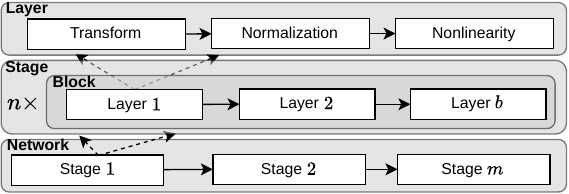}
    \caption{Network Architecture Layer Organization. A typical layer consists of a transform, normalization, and nonlinearity. A block contains $b$ layers, a stage repeats a block $n$ times, and the full network consists of $m$ stages.}
    \label{fig:comp_granul}
\end{figure}

\Cref{fig:comp_granul} illustrates how most modern architectures organize layers.
Each layer applies a linear transform, normalization, and introduces non-linearity with an activation function. Layers are grouped into blocks, which may be more complex, as shown here, such as the ResNet bottleneck~\cite{resnet} that uses two $1 \times 1$ convolutional layers to reduce the number of channels, before increasing them again. A stage consists of a sequence of repeated blocks. Finally, an architecture consists of at least one stage, and each stage may have a variable number of blocks. The difference between models of different sizes from the same architecture is typically their width and block ratios. For example, the block ratio in Swin-Tiny is 2:2:\textbf{6}:2 and in Swin-Base 2:2:\textbf{18}:2~\cite{swin}. 
Graph compilers can improve throughput by exploiting the repeated patterns present in such organizations and by providing specialized hardware modules for particular compositions (e.g., \texttt{Conv-BatchNorm-ReLU}). The following briefly summarizes software and hardware optimizations that compilers commonly use. 
\subsection{Software-Level Optimizations} \label{subsec:software_optimizations}
\subsubsection{Operator Fusion} \label{subsubsec:operator_fusion}
Operator fusion combines multiple operations, such as convolution and activation, into a single computational kernel. Without operator fusion, each operation would write intermediate results to memory and then read them back for the next operation. By fusing operations, the compiler generates a single kernel that executes all the fused operations sequentially within the same execution context. This eliminates redundant memory accesses and reduces the overhead of launching multiple kernels.
\subsubsection{Constant Folding} \label{subsubsec:constant_folding}
Constant folding identifies subgraphs where all inputs are constants and precomputes them at compile time. This reduces runtime computation by eliminating the need to compute results that do not depend on dynamic inputs repeatedly.
\subsubsection{Layout Transformation} \label{subsubsec:layout_transformation}
Different hardware architectures have specific data layout preferences for optimal performance. For example, 
NVIDIA performance guidance\footnote{https://docs.nvidia.com/deeplearning/performance/} often recommends NHWC over NCHW where possible, so TensorRT may internally transform tensors into HWC or other optimized formats.
Layout transformations reorganize tensor data into these preferred formats during compilation. These transformations ensure that memory accesses are coalesced and aligned with hardware requirements, improving throughput.
\subsection{Hardware and Kernel-Level Optimizations} \label{subsec:hardware_optimizations}
\subsubsection{Kernel Fusion} \label{subsubsec_kernel_fusion}
Kernel fusion is similar to operator fusion, but at the kernel level. Similar to operator fusion, kernel fusion combines multiple operations into one kernel execution to reduce kernel launch overheads.  However, kernel fusion operates at a lower level and can merge operations with finer granularity.  
\subsubsection{Memory Latency Hiding} \label{subsubsec:memory_latency_hiding}
Memory latency hiding overlaps computation with data transfers using asynchronous execution techniques. Specifically, by overlapping data movement, such as transfers between global memory and shared memory, with computation, compiler-generated schedules can hide part of the memory access latency. These schedules coordinate asynchronous memory transfers and thread execution so that available work can continue while other operations wait for data.
For example, in matrix multiplication on GPUs, while one block of threads computes partial results using data already loaded into shared memory, another block asynchronously loads the next set of data from global memory. 
\subsubsection{Sparse Computation} \label{subsubsec:sparse_computation}
Sparse computation exploits sparsity in weights or activations.
It leverages tensor sparsity patterns (e.g., weights with many zero values) to skip unnecessary calculations and reduce storage requirements and memory use. In particular, specialized sparse matrix formats like Compressed Sparse Row or Block Sparse Row store only non-zero elements efficiently. Hardware accelerators often include optimized sparse matrix multiplication routines that exploit these formats.
For example, consider a sparse neural network where 70\% of weights are zero due to unstructured pruning. Instead of performing dense matrix multiplication on all elements, sparse matrix multiplication algorithms process only non-zero elements stored in CSR format. Then, multiplying an input vector with a sparse weight matrix skips zero-weighted connections, reducing compute and memory usage.
\section{Graph Compiler-guided Solution Approach} \label{sec:framework}
We implement \textit{NGraphBench}, a library that permits quick, automated, empirical evaluation of graph compilers in a heterogeneous cluster. However, the focus of the work is not the implementation details of the library, and we only mention high-level details for evaluation transparency in \Cref{sec:eval}. Instead, the focus is on effectively utilizing empirical compiler benchmark results to iteratively conceive and refine ML methods, with a clear application focus on edge-cloud systems. 
\subsection{NGraphBench Library} \label{subsec:library}
\textit{NGraphBench} exposes a uniform interface for accessing and integrating graph compiler APIs. Users can provide their models in ONNX or native PyTorch and configure experiments, such as compiler-device pairs, compiler flags, repetitions, and model initialization parameters.
\begin{figure}[htb]
    \centering
    \includegraphics[width=\columnwidth]{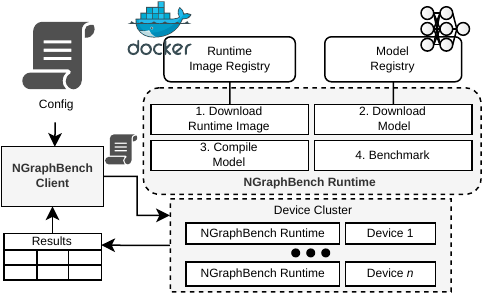}
    \caption{NGraphBench high-level flow. The client coordinates the experiments for participating devices in a cluster. Each device may evaluate multiple compilers.}
    \label{fig:lib_flow}
\end{figure}
\Cref{fig:lib_flow} illustrates the high-level application flow. The client will deploy the benchmarking application to the requested devices in the cluster. Crucially, the compilation is done \textit{locally} on the target devices, i.e., we are \textit{not} using hardware simulators, which are likely to result in worse optimizations. Each device may benchmark multiple compilers, and will persist results in predefined checkpoints periodically (e.g., to resume on a crash). After benchmarking, the devices will report the results to the client. Once all devices have reported their results, the client will tear down the benchmarking environments and terminate the application. 
\subsection{Pragmatic Research Design for Practical Systems} \label{sec:solapproach} 
\begin{figure}[htb]
    \centering
    \includegraphics[width=1\columnwidth]{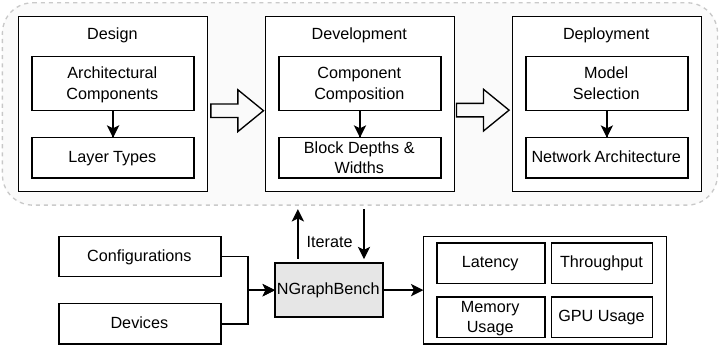}
    \caption{Iterative refinement guided by empirical analysis}
    \label{fig:design_Framework}
\end{figure}

The disconnect between academic research and practical deployment is particularly problematic when optimizing neural networks for heterogeneous hardware. While novel architectures may excel in controlled benchmarks, their performance can vary dramatically when deployed with different graph compilers across diverse hardware. We propose a methodological framework that incorporates compiler effects throughout the research process, as illustrated in \Cref{fig:design_Framework}.

This framework divides the research process into three phases, each integrating compiler optimization considerations:
\subsubsection{Design Phase} 
The design phase prioritizes architectural choices with widespread hardware and compiler support. Using our work in \cite{fool} as a case study, we selected variational compression methods for orbital edge computing applications where processing must occur within finite time windows. That case selected encoder components with broad compiler support to ensure results generalized across vendors; the present framework formalizes this practice and quantifies where compiler effects change design trade-offs. Rather than optimizing for theoretical metrics like the number of Multiply-And-Accumulate (MAC) operations or parameter counts, we introduced the \textit{Transfer Cost Reduction per Second (TCR/s)} metric to balance compression efficiency against computational throughput. This approach enabled the evaluation of different architectural paradigms (convolutional vs. transformer-based) against practical deployment metrics.
\subsubsection{Development Phase} 
During development, researchers must examine how block compositions affect model performance and compiler-optimized throughput. Our analysis revealed that increasing model depth often yields disproportionate throughput gains when target hardware incorporates vendor-specific compiler support. Similarly, compiler optimizations can effectively mitigate width adjustments that should reduce throughput. Researchers can identify viable block compositions and configurations that maximize performance within deployment constraints through systematic evaluation with graph compilers.
\subsubsection{Deployment Phase} 
The final phase involves comprehensive testing on target hardware with appropriate compiler optimizations. In our case study, without graph compiler optimization, increased model width significantly deteriorated batch parallelization efficiency, contradicting development-phase expectations. This resulted in selecting marginally smaller models for constrained devices despite 30\% worse compression performance. The cause was what we refer to as the \textit{batch-width scaling friction}, which we formalize in Section~\ref{subsubsec:bsrfric}. By slightly increasing the convolutional channels (i.e., the width), the TCR/s decreased significantly due to reduced processing throughput.
Such counterintuitive outcomes highlight the critical importance of evaluating compiler-hardware interactions throughout the research process.
As a concrete example, consider a neural compression model for edge inference under a finite processing window. In the design phase, the researcher first compares candidate encoder families, such as convolutional and transformer-based components, not only by parameter count or MACs, but by deployment-oriented metrics such as TCR/s after compilation. This may favor components with broad compiler support even when their theoretical complexity is not minimal. In the development phase, the researcher then varies the block composition, depth, and width of the selected family. Compiler feedback may show that adding repeated blocks can preserve or even improve compiled throughput because the compiler can exploit regular operator patterns, whereas increasing width may improve compression quality but reduce batch-scaling efficiency. In the deployment phase, the remaining candidates are compiled and benchmarked on the target devices under the intended batch sizes and runtime conditions. If the wider model achieves better compression but loses enough throughput due to batch-width scaling friction, the final deployment choice may be a smaller model with weaker compression quality but higher end-to-end TCR/s. This example illustrates how the framework turns compiler measurements into iterative design feedback rather than treating compilation as a final post-processing step.

In short, our methodology aims to bridge the gap between academic innovation and practical deployment by integrating graph compiler considerations into each research phase.
While this approach requires additional empirical testing, it ensures that reported performance improvements generalize across deployment scenarios, ultimately producing more valuable contributions for practitioners working with heterogeneous hardware environments.
\section{Evaluation} \label{sec:eval}
The evaluation is motivated by the three phases of our compiler-guided framework, as described in \Cref{sec:solapproach}. We examine: (1) differential compiler support across architectural styles to inform design decisions in \Cref{plot:throughput_by_device}; (2) depth scaling and batch-width friction mitigation effects to guide component composition in \Cref{subsec:depthscaling} and \Cref{subsec:frictionmitigation}; and (3) analyze graph compiler effects on resource usage to reason about unexpected throughput gains or losses, such as from interference from concurrent tasks in \Cref{subsec:loadreduction}.  
We restrict the empirical study to vision models to control confounders and isolate compiler effects, and our block-level experiments cover convolutional and attention-based compositions. Consequently, the specific throughput rankings and scaling patterns reported in this section should be interpreted as evidence for the evaluated model, compiler, and hardware configurations rather than as exhaustive claims across neural network domains. The proposed methodology is not vision-specific: it applies to systems that rely on ANNs for function approximation and whose deployed performance depends on compiler transformations, hardware kernels, batching behavior, and resource contention. However, when instantiated in other domains, including NLP and Deep Reinforcement Learning, workload-specific properties may alter opportunities for fusion, memory access patterns, control flow, and batch-scaling behavior. Examples include variable-length sequences, padding or bucketing strategies, autoregressive decoding, graph-structured inputs, multimodal pipelines, environment interaction loops, or recurrent computation. Such characteristics should therefore be evaluated with the same compiler-aware methodology before drawing deployment-specific conclusions. The experimentation framework required to reproduce our results is available as an open-source repository.\footnote{https://github.com/rezafuru/Graph-Compiler-Benchmarking}
\subsection{Methodology \& Experiment Design} 
We include TensorRT and OpenVINO to represent vendor-specific compilers and Apache TVM to represent vendor-agnostic compilers with hardware-level optimizations. The ONNX and TorchScript runtimes represent a software-level optimization approach. The evaluation exclusively focuses on applying graph compilers without fundamentally altering prediction behavior, i.e., it does \textit{not} consider quantization and other model compression methods.
For experiments with a relative measure that relies on a baseline (e.g., speedup factors, BSR), we use the PyTorch dynamic computational graph and refer to it as the \textit{identity}.
We repeat each experiment 100 times and report the average along with its standard deviation. Compilation is performed on the \textit{native hardware} without hardware simulators. The experiments are performed end-to-end by deploying them on a physical testbed cluster using the NGraphBench library (\Cref{sec:framework}). The following details the testbed and configurations to facilitate reproducibility. We emphasize that the NGraphBench library is exclusively for convenience and is not required to reproduce our results. 
\subsubsection{Testbed} \label{subsubsec:testbed}
We implement a physical testbed with relevant specifications summarized in \Cref{tab:testbed_hw}.
\begin{table}[htb]
\caption{Testbed Device Specifications}
\label{tab:testbed_hw}
\resizebox{\columnwidth}{!}{%
\begin{tabular}{lrr}
\hline
\multicolumn{1}{c}{Device} & \multicolumn{1}{c}{CPU} & \multicolumn{1}{c}{GPU} \\ \hline
\rowcolor[HTML]{EFEFEF} 
Server 1 & 8x Ryzen 5700G @ 3.80 GHz (x86) & RTX 4070 \\
Server 2 & 8x Xeon Skylake @ 3.0 GHz (x86) &  N/A \\
\rowcolor[HTML]{EFEFEF} 
Orin Nano  & 6x Cortex-A78 @ 2.0 GHz (ARM) & Amp. 512 CC 16 TC \\ \hline
\end{tabular}%
}
\end{table}

\noindent
For clarity, we will refer to Server 1 and Server 2 as “GPU” and “Xeon” respectively, i.e., the chip on which we compile and run the neural network.
The Orin Nano uses \texttt{JetPack 6.2}, which is based on \texttt{Ubuntu 22.04}. Hence, the other devices use \texttt{Ubuntu 22.04 LTS} with Linux kernel version 5.15. We prioritize consistency over using the latest versions. 
\Cref{tab:library_versions} reports the oldest versions installed on the devices. 
\begin{table}[htb]
\centering
\caption{Library Versions}
\label{tab:library_versions}
\resizebox{\columnwidth}{!}{%
\begin{subtable}[t]{0.5\columnwidth}
\centering
\begin{tabular}{lr}
Library     & Version   \\ \hline
\rowcolor[HTML]{EFEFEF} 
ONNX Runtime & 1.19.2    \\
TensorRT    & 10.4.0    \\
\rowcolor[HTML]{EFEFEF} 
Apache TVM   & 0.18.dev0 \\
OpenVINO    & 2024.3.0 
\end{tabular}
\end{subtable}%
\hfill
\begin{subtable}[t]{0.5\columnwidth}
\centering
\begin{tabular}{lr}
Library & Version \\ \hline
PyTorch & 2.4.1   \\
\rowcolor[HTML]{EFEFEF} 
CUDA    & 12.5    \\
cuDNN   & 9.3.0   \\
\rowcolor[HTML]{EFEFEF} 
timm    & 1.0.15
\end{tabular}
\end{subtable}%
}
\end{table}

\subsubsection{Compiler Configurations}
Except for Apache TVM, we use intuitive default configurations for graph compilers (e.g., optimize for throughput instead of latency in OpenVINO when the evaluation criterion is throughput). To remain vendor-agnostic, TVM takes a fundamentally different approach to optimization than vendor-specific compilers. 
TVM can fuse arbitrary patterns and support new operations, if it can find them~\cite{tvm}. Vendor-specific frameworks compile fast, but are limited to pre-defined fusion patterns or operations. 
TVM’s tuning involves running multiple candidate kernels on hardware or a simulator to measure performance, yielding highly optimized code that can match or exceed vendor libraries. 
The caveat is that TVM traverses an exponentially scaling search space, such that finding an optimized schedule/kernel configuration for a single experiment may take weeks or months. 
Therefore, we cap the number of trials at 1500 with early stopping after 150 trials using the \texttt{xgb} tuner, as we empirically determined on a subset of models that increasing the number of trials beyond 1500 yields diminishing gains. Moreover, we only apply TVM to the off-the-shelf models on the \textit{native hardware} and omit it from the block-level evaluation due to time constraints. 
These compile times reflect the combinatorial growth of the search space rather than a lack of generality. The weaker results on hybrid designs such as ConvNeXt may stem from the difficulty of finding good schedules in a large search space, which is consistent with the compile-time figures in \Cref{tab:compile_times}.
\begin{table}[htb]
\centering
\caption{Contrasting Compile Times in Seconds}
\label{tab:compile_times}
\resizebox{\columnwidth}{!}{%
\begin{tabular}{lcrrrr}
Model &
  Batch Size &
  \multicolumn{2}{r}{\textbf{Intel Xeon}} &
  \multicolumn{2}{r}{\textbf{GeForce RTX}} \\ \cline{3-6} 
 &
   &
  OpenVINO &
  TVM &
  TensorRT &
  TVM \\ \hline
ResNet-101 &
  \cellcolor[HTML]{EFEFEF}1 &
  \cellcolor[HTML]{EFEFEF}2.249 &
  \cellcolor[HTML]{EFEFEF}18,022.663 &
  \cellcolor[HTML]{EFEFEF}11.863 &
  \cellcolor[HTML]{EFEFEF}57,307.949 \\
 &
  32 &
  4.080 &
  28,327.363 &
  16.089 &
  58,466.694 \\ \hline
 &
  \cellcolor[HTML]{EFEFEF}1 &
  \cellcolor[HTML]{EFEFEF}3.109 &
  \cellcolor[HTML]{EFEFEF}36,241.423 &
  \cellcolor[HTML]{EFEFEF}53.332 &
  \cellcolor[HTML]{EFEFEF}69,364.992 \\
\multirow{-2}{*}{EfficientNet-B5} &
  32 &
  5.646 &
  49,765.446 &
  68.360 &
  61,700.123 \\ \hline
 &
  \cellcolor[HTML]{EFEFEF}1 &
  \cellcolor[HTML]{EFEFEF}4.830 &
  \cellcolor[HTML]{EFEFEF}36,431.073 &
  \cellcolor[HTML]{EFEFEF}10.004 &
  \cellcolor[HTML]{EFEFEF}16,533.937 \\
\multirow{-2}{*}{ConvNeXt-Base} &
  32 &
  6.763 &
  68,943.533 &
  19.394 &
  18,404.425 \\ \hline
 &
  \cellcolor[HTML]{EFEFEF}1 &
  \cellcolor[HTML]{EFEFEF}4.764 &
  \cellcolor[HTML]{EFEFEF}11,464.397 &
  \cellcolor[HTML]{EFEFEF}6.320 &
  \cellcolor[HTML]{EFEFEF}5,230.073 \\
\multirow{-2}{*}{DeiT-Base} &
  32 &
  6.264 &
  36,199.033 &
  13.827 &
  6,157.337 \\ \hline
 &
  \cellcolor[HTML]{EFEFEF}1 &
  \cellcolor[HTML]{EFEFEF}9.338 &
  \cellcolor[HTML]{EFEFEF}88,095.557 &
  \cellcolor[HTML]{EFEFEF}17.545 &
  \cellcolor[HTML]{EFEFEF}27,378.831 \\
\multirow{-2}{*}{Swin-Base} &
  32 &
  12.628 &
  134,262.784 &
  29.204 &
  14,287.950 \\ \hline
\end{tabular}%
}
\end{table}
\subsubsection{Network Architecture \& Layer Composition} \label{subsubsec:modelspecs}
We perform experiments on off-the-shelf architectures and more fine-grained blocks. Evaluating widespread models yields general insights, such as whether vendors favor a particular architectural style. 
\Cref{tab:archspecs} summarizes the architecture specifications. 

\begin{table}[htb]
\centering
\caption{Network Architecture Specifications}
\label{tab:archspecs}
\begin{tabular}{llrr}
Architecture    & Style         & Parameters & MACs          \\ \hline
\rowcolor[HTML]{EFEFEF} 
ResNet-18       & Convolutional & 11,689,512 & 1,814,083,944 \\
ResNet-50       & Convolutional & 25,557,032 & 4,089,238,376 \\
\rowcolor[HTML]{EFEFEF} 
ResNet-101      & Convolutional & 44,549,160 & 7,801,511,784 \\ \hline
EfficientNet-B3 & Convolutional & 12,233,232 & 962,729,320   \\
\rowcolor[HTML]{EFEFEF} 
EfficientNet-B4 & Convolutional & 19,341,616 & 1,503,740,472 \\
EfficientNet-B5 & Convolutional & 30,389,784 & 2,356,534,504 \\ \hline
\rowcolor[HTML]{EFEFEF} 
DeiT-Small      & Transformer   & 22,059,496 & 79,557,352    \\
DeiT-Medium     & Transformer   & 38,849,512 & 115,513,320   \\
\rowcolor[HTML]{EFEFEF} 
DeiT-Base       & Transformer   & 86,585,320 & 201,581,032   \\ \hline
Swin-Tiny       & Transformer   & 28,328,674 & 52,152,040    \\
\rowcolor[HTML]{EFEFEF} 
Swin-Small      & Transformer   & 49,737,298 & 66,312,424    \\
Swin-Base       & Transformer   & 71,125,762 & 94,739,176    \\ \hline
\rowcolor[HTML]{EFEFEF} 
ConvNeXt-Tiny   & Hybrid        & 28,589,128 & 322,371,592   \\
ConvNeXt-Small  & Hybrid        & 50,223,688 & 411,391,240   \\
\rowcolor[HTML]{EFEFEF} 
ConvNeXt-Base   & Hybrid        & 88,591,464 & 646,530,408   \\ \hline
\end{tabular}%
\end{table}
We use the \texttt{timm}~\cite{rw2019timm} library to access consistent off-the-shelf implementations and compute parameter and MAC counts with torchinfo\footnote{https://pypi.org/project/torchinfo/} under the same input configuration. Because MAC-counting conventions differ across tools and publications, we use these values primarily to compare relative model complexity within a single measurement pipeline rather than to reproduce the canonical complexity values reported in the original publications.
We consider five architectural families and three consecutively increasing model sizes per family. We include two convolutional-based (ResNets~\cite{resnet}, EfficientNets~\cite{efficientnet}) and two transformer-based (Swins~\cite{swin}, DeiTs~\cite{deit}) architectures. Additionally, we include ConvNeXts~\cite{convnext} as a hybrid approach, which is a convolutional-based model with design principles from transformers.
\Cref{tab:blockspecs} summarizes the per-block specifications.
\begin{table}[htb]
\centering
\caption{Select Per-Block Specifications}
\label{tab:blockspecs}
\resizebox{\columnwidth}{!}{%
\begin{tabular}{rrrr}
\multicolumn{2}{r}{\textbf{Convolutional}} & \multicolumn{2}{r}{\textbf{Multi-Head Attention}}                 \\ \hline
Channels    & Params Per Block    & Embedding Dimensions & Params Per Block \\ \hline
\rowcolor[HTML]{EFEFEF} 
64          & 37,056              & 128                  & 66,048           \\
96          & 83,232              & 256                  & 263,168          \\
\rowcolor[HTML]{EFEFEF} 
128         & 147,840             & 384                  & 591,360          \\
256         & 590,592             & 512                  & 1,050,624        \\ \hline
\end{tabular}%
}
\end{table}

\noindent
Blocks enable a more targeted evaluation of depth (i.e., investigating optimization as we stack repeated blocks) and width, and reduce noise from certain implementation quirks or other factors that affect compiler efficacy. 
The multi-head attention (MHA) block uses ReLU nonlinearity. We use channels in convolutional blocks and embedding dimensions for MHA blocks to parameterize block widths when investigating batch-width friction. We found that varying kernel and input sizes have a similar effect on batch-width friction as increasing the number of channels. To simplify, we only report results with the kernel size fixed at $3 \times 3$ and the input size set to $3 \times 224 \times 224$. In MHA block experiments, we fix the sequence length to ten for the input tensor and consider the embedding dimensions to parameterize the block width.
The block-level configurations are not intended to define parameter- or FLOP-equivalent convolutional and MHA architectures. Instead, they provide representative operating points for analyzing compiler behavior under controlled changes in width, depth, and batch size. \Cref{plot:speedup_throughput_inc_stacks_rel_combined} reports the broader width and depth sweep, while \Cref{tab:depth_scaling_conv} and \Cref{tab:depth_scaling_mha} summarize selected intermediate batch-size configurations through slope and retention metrics. This separation keeps the qualitative trend visible across the full sweep and the quantitative comparison focused on operating points large enough for batching effects to dominate launch overheads, but not so large that the largest configurations dominate the results through resource saturation. Convolutional width is parameterized by the number of channels, while MHA width is parameterized by the embedding dimension; these parameters affect compute and memory pressure differently, but both serve as the primary width controls within their respective block families. Thus, the comparison should be interpreted as a normalized compiler-behavior comparison across representative operating points, rather than as a claim of architectural equivalence between convolutional and attention blocks.
\subsubsection{Measuring Batch Parallelization} \label{subsubsec:measuringbatchpar}
We can measure the Relative Throughput Rate (RTR) as
\[
    RTR_c(b) = \frac{T_c(b)}{T_c(1)}
\]

\noindent where $T_c(b)$ is the throughput in samples per second for batch size $b$ when compiler $c$ is used, for a given width ($w$) configuration. For notation simplicity, we omit the width parameter from the expressions of this section when it is fixed. 
The RTR quantifies the unnormalized parallelization rate. When scaling is perfect, the throughput scales linearly with the batch size. Note that once RTR drops below 1, batching reduces absolute throughput. Moreover, ideal scaling is not expected in practice for larger batch sizes, so we will visualize the decay in batch scaling efficiency with the Absolute Scaling Efficiency (ASE) measure:
\[
    ASE_c(b) = \frac{T_c(b)}{b \cdot T_c(1)}.
\]

\noindent The ASE normalizes the RTR, so any reduction from a perfect parallelization rate directly indicates reduced scaling efficiency. For example, when batch parallelization is ideal, the ASE remains consistently at $100\%$, irrespective of the batch size.
Conversely, decreasing ASE implies diminishing returns from increasing the batch size.
As discussed in \Cref{sec:solapproach}, it is interesting to see whether compilers can mitigate the decay, i.e., maintain scaling efficiency at higher batch sizes.
\noindent We measure this with the Batch Scaling Resilience (BSR) metric as follows:
\[
BSR_c(b) = \frac{ASE_c(b)}{ASE_{identity}(b)}.
\]

\noindent The BSR is a relative measure that can quantify, without eyeballing, the improvement in mitigating friction compared to a baseline compiler. Compilers with BSR values $>1$ consistently across varying configurations demonstrate that they can improve the batch scaling efficiency for target hardware.  

We emphasize that ASE and BSR are analytical instruments that quantify compiler-induced scaling behavior; they are not conceptual innovations but make observed patterns measurable and comparable across settings. In particular, measuring BSR as we increase the batch size for different width configurations can determine whether a compiler can alleviate the scaling friction, which we elaborate on in the following. 
\subsubsection{Batch-Width Scaling Friction} \label{subsubsec:bsrfric}
The \textit{batch-width scaling friction} describes the joint effects of increasing model width and batch size parameters on scaling efficiency that significantly impact systems that prioritize processing throughput, such as in \cite{fool}.
To empirically assess the efficacy of compilers to mitigate the effect, we provide a minimally formal definition. 
\begin{definition}{(Batch-Width Scaling Friction)}
Batch-width scaling friction is present when, for fixed $b$, $ASE_c(b,w_2) < ASE_c(b,w_1)$ with widths $w_2 > w_1$ in the operating range, which indicates that increasing width reduces parallelization efficiency.
\end{definition}
\subsection{Differential Compiler Support Across Architectural Styles} \label{subsec:throughput}
\begin{figure}[htb]
    \centering
    \begin{subfigure}[b]{\columnwidth}
        \centering
        \includegraphics[width=\textwidth]{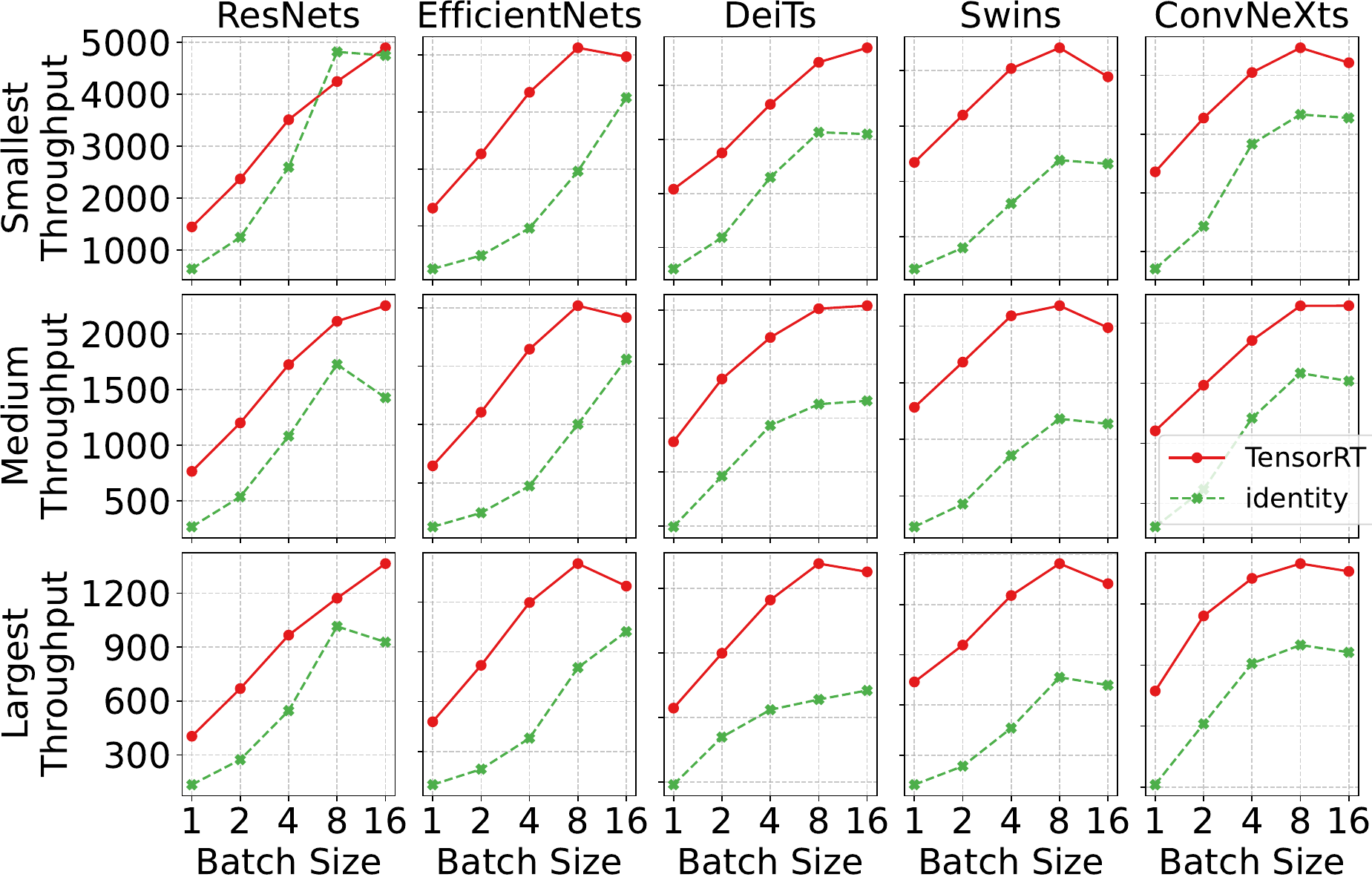}
        \caption{Device = GPU}
        \label{plot:throughput_by_network_abs_gpu}
    \end{subfigure}
    
    \vspace{1px} 
    
    \begin{subfigure}[b]{\columnwidth}
        \centering
        \includegraphics[width=\textwidth]{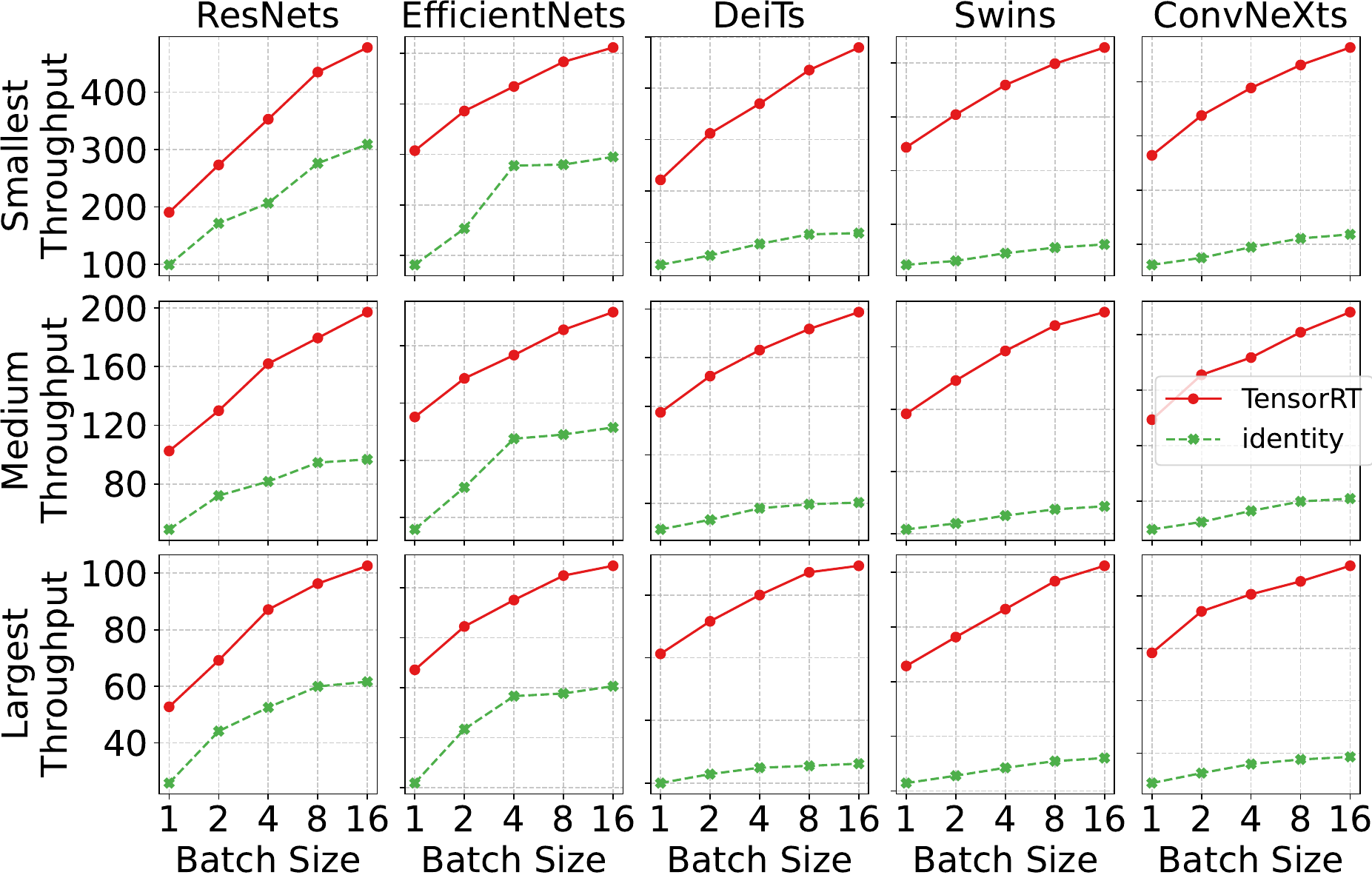}
        \caption{Device = Orin}
        \label{plot:throughput_by_network_abs_orin}
    \end{subfigure}
    
    \vspace{1px} 
    
    \begin{subfigure}[b]{\columnwidth}
        \centering
        \includegraphics[width=\textwidth]{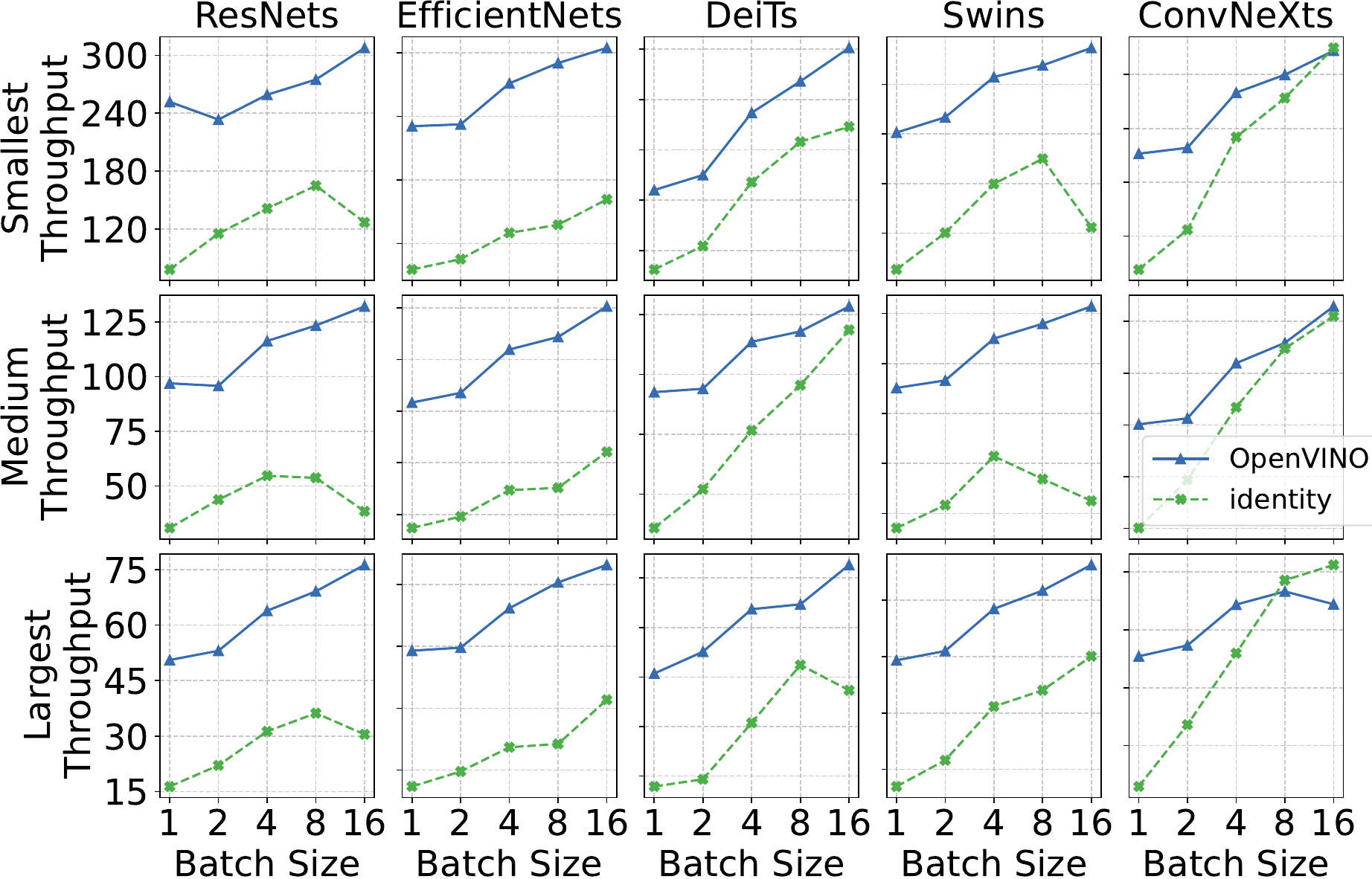}
        \caption{Device = Xeon}
        \label{plot:throughput_by_network_abs_xeon}
    \end{subfigure}
    
    \caption{
    Absolute throughput in samples per second for vendor-specific compilers and the dynamic graph baseline. Each row is a model size within an architectural family. Compilation increases throughput in most configurations, with gains depending on device, model family, and batch size.
     }
    \label{plot:throughput_by_device}
\end{figure}

\Cref{plot:throughput_by_device} compares the absolute throughput of the vendor-specific compilers with the dynamic uncompiled graphs. Each row corresponds to a model size, and each column to an architectural family (e.g., smallest for ResNets is ResNet-18). 
The y-axis scaling is non-uniform to highlight the strong relationship between compiler efficacy and architectural style. Notice that even if the absolute throughput across model sizes is offset, the throughput scaling as we increase the batch size is strikingly similar within a family.

The caveat is that the varying device capacities
obfuscate the results, making it challenging to assess compilation efficacy.
Hence, we report relative values for the remainder of the evaluation but summarize partially aggregated measurements using absolute values in \Cref{tab:network_results_summary}.
\begin{table*}[htb]
\centering
\caption{Aggregated Results by Architectural Family}
\label{tab:network_results_summary}
\resizebox{2\columnwidth}{!}{%
\begin{tabular}{lllrrrrrr}
 &
   &
   &
  \multicolumn{2}{r}{Batch Size (1--2)} &
  \multicolumn{2}{r}{Batch Size (4--8)} &
  \multicolumn{2}{r}{Batch Size (16--32)} \\ \cline{4-9} 
\multirow{-2}{*}{Family} &
  \multirow{-2}{*}{Compiler} &
  \multirow{-2}{*}{Device} &
  \multicolumn{1}{l}{Throughput [t/s] $\uparrow$} &
  \multicolumn{1}{l}{CPU Usage [\%] $\downarrow$} &
  \multicolumn{1}{l}{Throughput [t/s] $\uparrow$} &
  \multicolumn{1}{l}{CPU Usage [\%] $\downarrow$} &
  \multicolumn{1}{l}{Throughput [t/s] $\uparrow$} &
  \multicolumn{1}{l}{CPU Usage [\%] $\downarrow$} \\ \hline
 &
  \cellcolor[HTML]{EFEFEF}Identity &
  \cellcolor[HTML]{EFEFEF}GPU &
  \cellcolor[HTML]{EFEFEF}296.42 $\pm$ 3.15 &
  \cellcolor[HTML]{EFEFEF}6.24 $\pm$ 0.19 &
  \cellcolor[HTML]{EFEFEF}802.75$\pm$4.97 &
  \cellcolor[HTML]{EFEFEF}6.25$\pm$0.1 &
  \cellcolor[HTML]{EFEFEF}836.9$\pm$2.65 &
  \cellcolor[HTML]{EFEFEF}6.25$\pm$0.1 \\
 &
  Identity &
  Orin Nano &
  26.92 $\pm$ 0.16 &
  9.44 $\pm$ 1.32 &
  32.57$\pm$0.13 &
  1.49$\pm$2.76 &
  34.93$\pm$0.1 &
  1.49$\pm$2.76 \\
 &
  \cellcolor[HTML]{EFEFEF}Identity &
  \cellcolor[HTML]{EFEFEF}Xeon &
  \cellcolor[HTML]{EFEFEF}24.27 $\pm$ 2.1 &
  \cellcolor[HTML]{EFEFEF}98.22 $\pm$ 3.25 &
  \cellcolor[HTML]{EFEFEF}39.72$\pm$1.23 &
  \cellcolor[HTML]{EFEFEF}93.67$\pm$4.27 &
  \cellcolor[HTML]{EFEFEF}36.8$\pm$0.51 &
  \cellcolor[HTML]{EFEFEF}93.67$\pm$4.27 \\
 &
  OpenVINO &
  Xeon &
  35.4 $\pm$ 0.99 &
  99.66$\pm$0.91 &
  43.1$\pm$1.13 &
  95.43$\pm$2.28 &
  46.41$\pm$1.19 &
  95.43$\pm$2.28 \\
 &
  \cellcolor[HTML]{EFEFEF}TVM &
  \cellcolor[HTML]{EFEFEF}GPU &
  \cellcolor[HTML]{EFEFEF}189.6 $\pm$ 18.14 &
  \cellcolor[HTML]{EFEFEF}6.25$\pm$0.06 &
  \cellcolor[HTML]{EFEFEF}217.93$\pm$36.73 &
  \cellcolor[HTML]{EFEFEF}6.25$\pm$0.09 &
  \cellcolor[HTML]{EFEFEF}109.88$\pm$5.19 &
  \cellcolor[HTML]{EFEFEF}6.25$\pm$0.09 \\
 &
  TVM &
  Xeon &
  8.42 $\pm$ 0.13 &
  49.97$\pm$0.75 &
  12.51$\pm$0.26 &
  49.91$\pm$0.86 &
  13.76$\pm$0.22 &
  49.91$\pm$0.86 \\
 &
  \cellcolor[HTML]{EFEFEF}TensorRT &
  \cellcolor[HTML]{EFEFEF}GPU &
  \cellcolor[HTML]{EFEFEF}750.61 $\pm$ 6.18 &
  \cellcolor[HTML]{EFEFEF}6.24$\pm$0.07 &
  \cellcolor[HTML]{EFEFEF}1128.36$\pm$6.35 &
  \cellcolor[HTML]{EFEFEF}6.25$\pm$0.09 &
  \cellcolor[HTML]{EFEFEF}1094.74$\pm$4.48 &
  \cellcolor[HTML]{EFEFEF}6.25$\pm$0.09 \\
\multirow{-8}{*}{ConvNeXts} &
  TensorRT &
  Orin Nano &
  67.13 $\pm$ 0.52 &
  3.41$\pm$1.19 &
  83.74$\pm$0.54 &
  0.54$\pm$0.77 &
  93.04$\pm$0.39 &
  0.54$\pm$0.77 \\ \hline
 &
  \cellcolor[HTML]{EFEFEF}Identity &
  \cellcolor[HTML]{EFEFEF}GPU &
  \cellcolor[HTML]{EFEFEF}155.2$\pm$1.31 &
  \cellcolor[HTML]{EFEFEF}6.24$\pm$0.1 &
  \cellcolor[HTML]{EFEFEF}504.09$\pm$2.32 &
  \cellcolor[HTML]{EFEFEF}6.25$\pm$0.08 &
  \cellcolor[HTML]{EFEFEF}560.38$\pm$0.77 &
  \cellcolor[HTML]{EFEFEF}6.25$\pm$0.09 \\
 &
  Identity &
  Orin Nano &
  20.36$\pm$0.09 &
  10.97$\pm$1.47 &
  24.32$\pm$0.08 &
  4.86$\pm$4.03 &
  26.14$\pm$0.04 &
  2.13$\pm$4.12 \\
 &
  \cellcolor[HTML]{EFEFEF}Identity &
  \cellcolor[HTML]{EFEFEF}Xeon &
  \cellcolor[HTML]{EFEFEF}15.88$\pm$0.53 &
  \cellcolor[HTML]{EFEFEF}95.52$\pm$2.12 &
  \cellcolor[HTML]{EFEFEF}25.42$\pm$0.56 &
  \cellcolor[HTML]{EFEFEF}97.72$\pm$2.41 &
  \cellcolor[HTML]{EFEFEF}18.95$\pm$0.33 &
  \cellcolor[HTML]{EFEFEF}90.8$\pm$4.3 \\
 &
  OpenVINO &
  Xeon &
  33.14$\pm$1.16 &
  99.46$\pm$0.97 &
  40.52$\pm$0.86 &
  98.84$\pm$1.17 &
  43.43$\pm$0.76 &
  97.33$\pm$2.84 \\
 &
  \cellcolor[HTML]{EFEFEF}TVM &
  \cellcolor[HTML]{EFEFEF}GPU &
  \cellcolor[HTML]{EFEFEF}102.52$\pm$7.4 &
  \cellcolor[HTML]{EFEFEF}6.24$\pm$0.04 &
  \cellcolor[HTML]{EFEFEF}124.08$\pm$17.16 &
  \cellcolor[HTML]{EFEFEF}6.27$\pm$0.14 &
  \cellcolor[HTML]{EFEFEF}77.05$\pm$1.63 &
  \cellcolor[HTML]{EFEFEF}6.26$\pm$0.15 \\
 &
  TVM &
  Xeon &
  6.61$\pm$0.11 &
  49.96$\pm$0.78 &
  9.03$\pm$0.18 &
  49.92$\pm$0.83 &
  9.72$\pm$0.14 &
  49.89$\pm$0.85 \\
 &
  \cellcolor[HTML]{EFEFEF}TensorRT &
  \cellcolor[HTML]{EFEFEF}GPU &
  \cellcolor[HTML]{EFEFEF}708.65$\pm$7.62 &
  \cellcolor[HTML]{EFEFEF}6.22$\pm$0.21 &
  \cellcolor[HTML]{EFEFEF}1073.46$\pm$6.04 &
  \cellcolor[HTML]{EFEFEF}6.24$\pm$0.17 &
  \cellcolor[HTML]{EFEFEF}924.43$\pm$3.14 &
  \cellcolor[HTML]{EFEFEF}6.25$\pm$0.1 \\
\multirow{-8}{*}{Swins} &
  TensorRT &
  Orin Nano &
  58.93$\pm$0.52 &
  1.95$\pm$0.78 &
  75.84$\pm$0.47 &
  0.78$\pm$0.82 &
  84.48$\pm$1.18 &
  0.46$\pm$0.96 \\ \hline
 &
  \cellcolor[HTML]{EFEFEF}Identity &
  \cellcolor[HTML]{EFEFEF}GPU &
  \cellcolor[HTML]{EFEFEF}410.58$\pm$5.18 &
  \cellcolor[HTML]{EFEFEF}6.24$\pm$0.09 &
  \cellcolor[HTML]{EFEFEF}918.43$\pm$4.37 &
  \cellcolor[HTML]{EFEFEF}6.26$\pm$0.15 &
  \cellcolor[HTML]{EFEFEF}1028.15$\pm$3.12 &
  \cellcolor[HTML]{EFEFEF}6.26$\pm$0.09 \\
 &
  Identity &
  Orin Nano &
  32.35$\pm$0.26 &
  8.98$\pm$1.24 &
  39.91$\pm$0.22 &
  3.65$\pm$1.72 &
  42.39$\pm$0.12 &
  1.02$\pm$1.79 \\
 &
  \cellcolor[HTML]{EFEFEF}Identity &
  \cellcolor[HTML]{EFEFEF}Xeon &
  \cellcolor[HTML]{EFEFEF}30.02$\pm$1.19 &
  \cellcolor[HTML]{EFEFEF}99.75$\pm$1.31 &
  \cellcolor[HTML]{EFEFEF}45.54$\pm$1.47 &
  \cellcolor[HTML]{EFEFEF}99.51$\pm$1.65 &
  \cellcolor[HTML]{EFEFEF}54.89$\pm$1.21 &
  \cellcolor[HTML]{EFEFEF}99.2$\pm$1.37 \\
 &
  OpenVINO &
  Xeon &
  44.79$\pm$1.28 &
  99.44$\pm$0.93 &
  56.62$\pm$1.95 &
  98.36$\pm$1.6 &
  63.5$\pm$1.43 &
  97.07$\pm$2.55 \\
 &
  \cellcolor[HTML]{EFEFEF}TVM &
  \cellcolor[HTML]{EFEFEF}GPU &
  \cellcolor[HTML]{EFEFEF}66.42$\pm$3.07 &
  \cellcolor[HTML]{EFEFEF}6.27$\pm$0.09 &
  \cellcolor[HTML]{EFEFEF}67.67$\pm$6.9 &
  \cellcolor[HTML]{EFEFEF}6.25$\pm$0.08 &
  \cellcolor[HTML]{EFEFEF}96.59$\pm$3.7 &
  \cellcolor[HTML]{EFEFEF}6.25$\pm$0.1 \\
 &
  TVM &
  Xeon &
  5.53$\pm$0.05 &
  49.94$\pm$0.84 &
  7.83$\pm$0.09 &
  49.94$\pm$0.88 &
  11.01$\pm$0.08 &
  49.86$\pm$0.83 \\
 &
  \cellcolor[HTML]{EFEFEF}TensorRT &
  \cellcolor[HTML]{EFEFEF}GPU &
  \cellcolor[HTML]{EFEFEF}894.98$\pm$9.53 &
  \cellcolor[HTML]{EFEFEF}6.31$\pm$0.22 &
  \cellcolor[HTML]{EFEFEF}1404.31$\pm$8.25 &
  \cellcolor[HTML]{EFEFEF}6.27$\pm$0.15 &
  \cellcolor[HTML]{EFEFEF}1487.99$\pm$9.53 &
  \cellcolor[HTML]{EFEFEF}6.25$\pm$0.09 \\
\multirow{-8}{*}{DeiTs} &
  TensorRT &
  Orin Nano &
  81.63$\pm$3.25 &
  2.14$\pm$0.92 &
  107.02$\pm$0.81 &
  0.88$\pm$0.79 &
  121.65$\pm$0.52 &
  0.45$\pm$0.76 \\ \hline
 &
  \cellcolor[HTML]{EFEFEF}Identity &
  \cellcolor[HTML]{EFEFEF}GPU &
  \cellcolor[HTML]{EFEFEF}149.66$\pm$1.17 &
  \cellcolor[HTML]{EFEFEF}6.26$\pm$0.08 &
  \cellcolor[HTML]{EFEFEF}604.34$\pm$6.82 &
  \cellcolor[HTML]{EFEFEF}6.24$\pm$0.09 &
  \cellcolor[HTML]{EFEFEF}1162.07$\pm$1.76 &
  \cellcolor[HTML]{EFEFEF}6.25$\pm$0.09 \\
 &
  Identity &
  Orin Nano &
  29.57$\pm$0.2 &
  16.43$\pm$0.61 &
  62.28$\pm$0.18 &
  9.71$\pm$1.82 &
  66.8$\pm$0.13 &
  4.97$\pm$4.98 \\
 &
  \cellcolor[HTML]{EFEFEF}Identity &
  \cellcolor[HTML]{EFEFEF}Xeon &
  \cellcolor[HTML]{EFEFEF}21.8$\pm$0.71 &
  \cellcolor[HTML]{EFEFEF}99.1$\pm$1.23 &
  \cellcolor[HTML]{EFEFEF}38.19$\pm$0.73 &
  \cellcolor[HTML]{EFEFEF}98.27$\pm$2.13 &
  \cellcolor[HTML]{EFEFEF}43.66$\pm$0.81 &
  \cellcolor[HTML]{EFEFEF}90.72$\pm$7.22 \\
 &
  OpenVINO &
  Xeon &
  84.98$\pm$2.63 &
  99.01$\pm$0.83 &
  110.48$\pm$2.86 &
  97.85$\pm$0.99 &
  122.77$\pm$2.68 &
  96.83$\pm$1.98 \\
 &
  \cellcolor[HTML]{EFEFEF}TVM &
  \cellcolor[HTML]{EFEFEF}GPU &
  \cellcolor[HTML]{EFEFEF}648.41$\pm$53.97 &
  \cellcolor[HTML]{EFEFEF}6.17$\pm$0.1 &
  \cellcolor[HTML]{EFEFEF}1261.1$\pm$146.94 &
  \cellcolor[HTML]{EFEFEF}6.25$\pm$0.15 &
  \cellcolor[HTML]{EFEFEF}1261.35$\pm$220.69 &
  \cellcolor[HTML]{EFEFEF}6.25$\pm$0.07 \\
 &
  TVM &
  Xeon &
  16.26$\pm$0.4 &
  49.99$\pm$0.76 &
  19.36$\pm$0.47 &
  49.94$\pm$0.8 &
  20.62$\pm$0.4 &
  49.94$\pm$0.86 \\
 &
  \cellcolor[HTML]{EFEFEF}TensorRT &
  \cellcolor[HTML]{EFEFEF}GPU &
  \cellcolor[HTML]{EFEFEF}711.79$\pm$5.13 &
  \cellcolor[HTML]{EFEFEF}6.24$\pm$0.11 &
  \cellcolor[HTML]{EFEFEF}1476.8$\pm$8.64 &
  \cellcolor[HTML]{EFEFEF}6.23$\pm$0.15 &
  \cellcolor[HTML]{EFEFEF}1470.06$\pm$7.4 &
  \cellcolor[HTML]{EFEFEF}6.25$\pm$0.15 \\
\multirow{-8}{*}{EfficientNets} &
  TensorRT &
  Orin Nano &
  79.4$\pm$0.73 &
  8.42$\pm$1.36 &
  104.69$\pm$0.35 &
  3.34$\pm$0.83 &
  116.84$\pm$0.37 &
  1.34$\pm$1.12 \\ \hline
 &
  \cellcolor[HTML]{EFEFEF}Identity &
  \cellcolor[HTML]{EFEFEF}GPU &
  \cellcolor[HTML]{EFEFEF}516.67$\pm$6.73 &
  \cellcolor[HTML]{EFEFEF}6.21$\pm$0.14 &
  \cellcolor[HTML]{EFEFEF}1047.32$\pm$14.09 &
  \cellcolor[HTML]{EFEFEF}6.24$\pm$0.09 &
  \cellcolor[HTML]{EFEFEF}2342.31$\pm$7.52 &
  \cellcolor[HTML]{EFEFEF}6.24$\pm$0.07 \\
 &
  Identity &
  Orin Nano &
  76.88$\pm$2.57 &
  13.55$\pm$0.83 &
  104.69$\pm$0.57 &
  11.13$\pm$0.78 &
  161.64$\pm$0.49 &
  3.21$\pm$2.52 \\
 &
  \cellcolor[HTML]{EFEFEF}Identity &
  \cellcolor[HTML]{EFEFEF}Xeon &
  \cellcolor[HTML]{EFEFEF}51.08$\pm$2.27 &
  \cellcolor[HTML]{EFEFEF}99.7$\pm$1.12 &
  \cellcolor[HTML]{EFEFEF}68.08$\pm$1.75 &
  \cellcolor[HTML]{EFEFEF}99.83$\pm$0.94 &
  \cellcolor[HTML]{EFEFEF}62.64$\pm$1.95 &
  \cellcolor[HTML]{EFEFEF}89.63$\pm$5.71 \\
 &
  OpenVINO &
  Xeon &
  130.26$\pm$6.67 &
  98.96$\pm$0.76 &
  136.94$\pm$5.80 &
  97.98$\pm$0.84 &
  172.72$\pm$4.6 &
  96.71$\pm$2.09 \\
 &
  \cellcolor[HTML]{EFEFEF}TVM &
  \cellcolor[HTML]{EFEFEF}GPU &
  \cellcolor[HTML]{EFEFEF}983.08$\pm$87.23 &
  \cellcolor[HTML]{EFEFEF}6.33$\pm$0.16 &
  \cellcolor[HTML]{EFEFEF}1678.78$\pm$29.22 &
  \cellcolor[HTML]{EFEFEF}6.32$\pm$0.17 &
  \cellcolor[HTML]{EFEFEF}1774.73$\pm$340.74 &
  \cellcolor[HTML]{EFEFEF}6.28$\pm$0.08 \\
 &
  TVM &
  Xeon &
  39.06$\pm$1.16 &
  50.04$\pm$0.68 &
  46.55$\pm$0.69 &
  49.99$\pm$0.72 &
  59.68$\pm$1.34 &
  49.87$\pm$0.85 \\
 &
  \cellcolor[HTML]{EFEFEF}TensorRT &
  \cellcolor[HTML]{EFEFEF}GPU &
  \cellcolor[HTML]{EFEFEF}1143.11$\pm$11.2 &
  \cellcolor[HTML]{EFEFEF}6.25$\pm$0.04 &
  \cellcolor[HTML]{EFEFEF}1740.11$\pm$14.18 &
  \cellcolor[HTML]{EFEFEF}6.26$\pm$0.14 &
  \cellcolor[HTML]{EFEFEF}2925.29$\pm$21.12 &
  \cellcolor[HTML]{EFEFEF}6.25$\pm$0.09 \\
\multirow{-8}{*}{ResNets} &
  TensorRT &
  Orin Nano &
  136.31$\pm$11.09 &
  4.41$\pm$1.03 &
  179.01$\pm$6.98 &
  3.45$\pm$1.14 &
  274.97$\pm$1.71 &
  1.18$\pm$0.82
\end{tabular}%
}
\end{table*}

\begin{figure}[htb]
    \centering
    \includegraphics[width=\columnwidth]{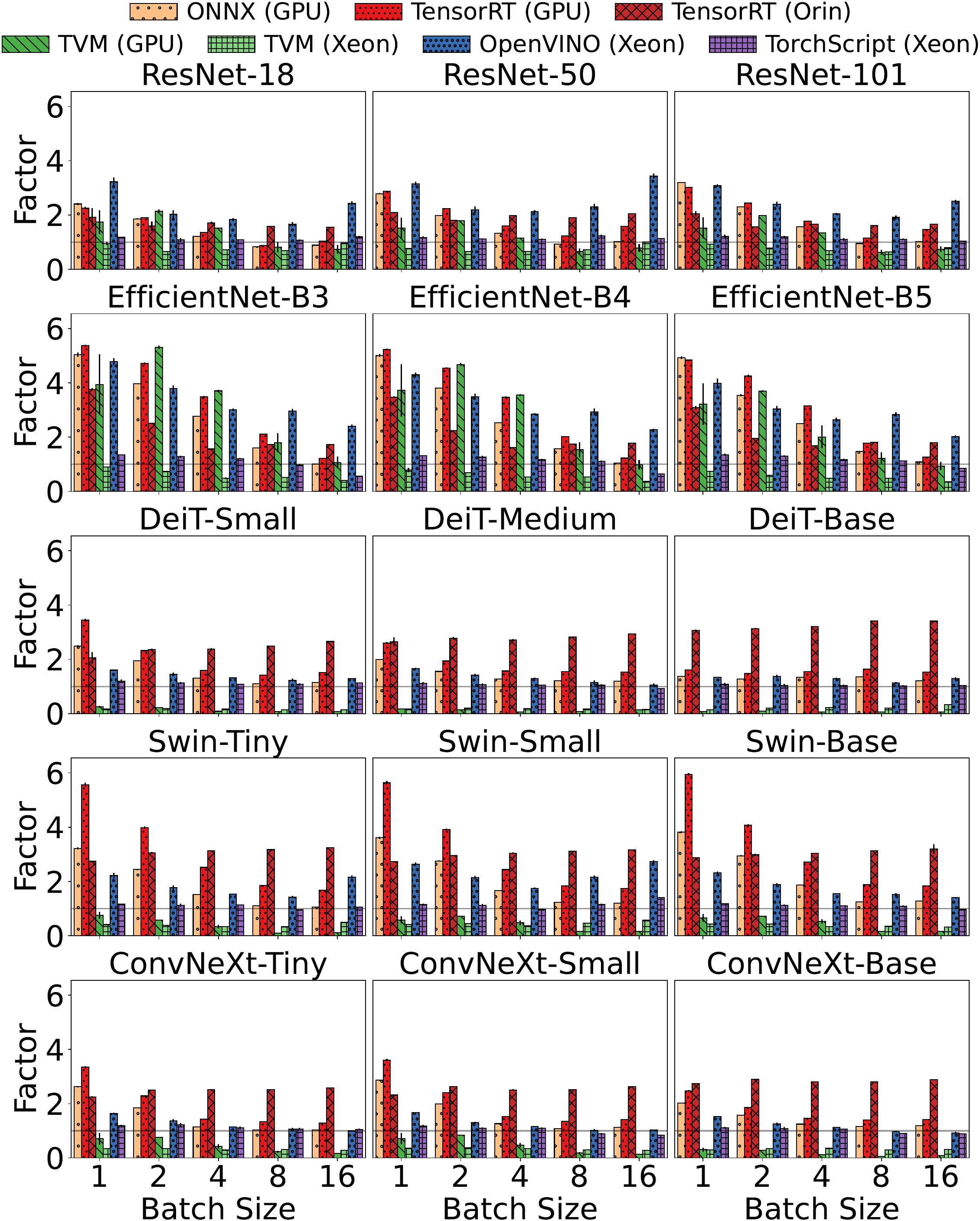}
    \caption{
    Relative speed-up over the dynamic graph baseline. Values above 1 indicate higher throughput for the same model, device, and batch size. Convolutional architectures show broader compiler support, while transformer-based architectures depend more on the compiler-device pair.
    }
    \label{plot:speedup_throughput_inc_network_rel}
\end{figure}

\Cref{plot:speedup_throughput_inc_network_rel} plots the throughput multiplier relative to the dynamic graph for the five architectural families on different device-compiler pairs. 
The results reveal strikingly distinct performance dependency patterns on both the batch size \textit{and} the underlying neural network architecture.

Interestingly, despite being advertised as a ``convolutional architecture,'' ConvNeXt behaves similarly to transformers in performance across all compilers. 
At small batches ($\leq$2), several compiler-device pairs provide 2-6$\times$ speed-ups by eliminating Python dispatch overhead and enabling operator fusion.
These advantages diminish as the batch size increases, with only vendor-specific solutions (TensorRT for GPU and OpenVINO for CPU) maintaining consistent advantages at a batch size of 16. 

In summary, architecture significantly influences compiler efficacy. 
Traditional convolutional networks benefit substantially from most compiler-device pairs at small batches, while transformer-based models show minimal improvement with TVM, moderate gains with ONNX, and substantial acceleration mainly with vendor-specific tools.
While TVM demonstrates substantial performance gains for convolutional architectures, its gains are substantially weaker for transformer-based architectures and ConvNeXt. We interpret the weaker TVM results on ConvNeXt as a consequence of search-based optimization over a larger schedule space for hybrid blocks [22], which is consistent with the compile-time profile in Table III. The architectural hybridization of ConvNeXt may increase the difficulty of identifying effective schedules and fusion candidates under the fixed tuning budget used in this study. Conversely, vendor-specific compilers may benefit from manually optimized support for widely used architectures. We do not claim this as a root-cause analysis of the generated kernels, but as an interpretation consistent with the measured throughput and compile-time results.
%
Since the dynamic computational graph is executed sequentially in the Python runtime on the Xeon CPU, we also include TorchScript as an additional baseline. TorchScript can exploit the multiple cores on the CPU with intra-op parallelism. However, compared to OpenVINO, which achieves up to 5-6 times higher throughput, TorchScript only marginally but consistently improves throughput, which is expected considering that it primarily applies rudimentary software-level optimizations.

\bboxed{
The results show distinct performance patterns across compilers, hardware platforms, and architectures. 
Vendor-specific compilers maintain advantages through targeted optimization for popular architectures, while automated tuning approaches struggle with hybrid designs. 
}
\subsection{Exploiting Repeated Patterns from Depth Scaling} \label{subsec:depthscaling}
\begin{figure}[htb]
    \centering 
    \subfloat[Batch Size = 2\label{plot:speedup_throughput_inc_stacks_rel_bs1}]{%
        \includegraphics[width=\columnwidth]{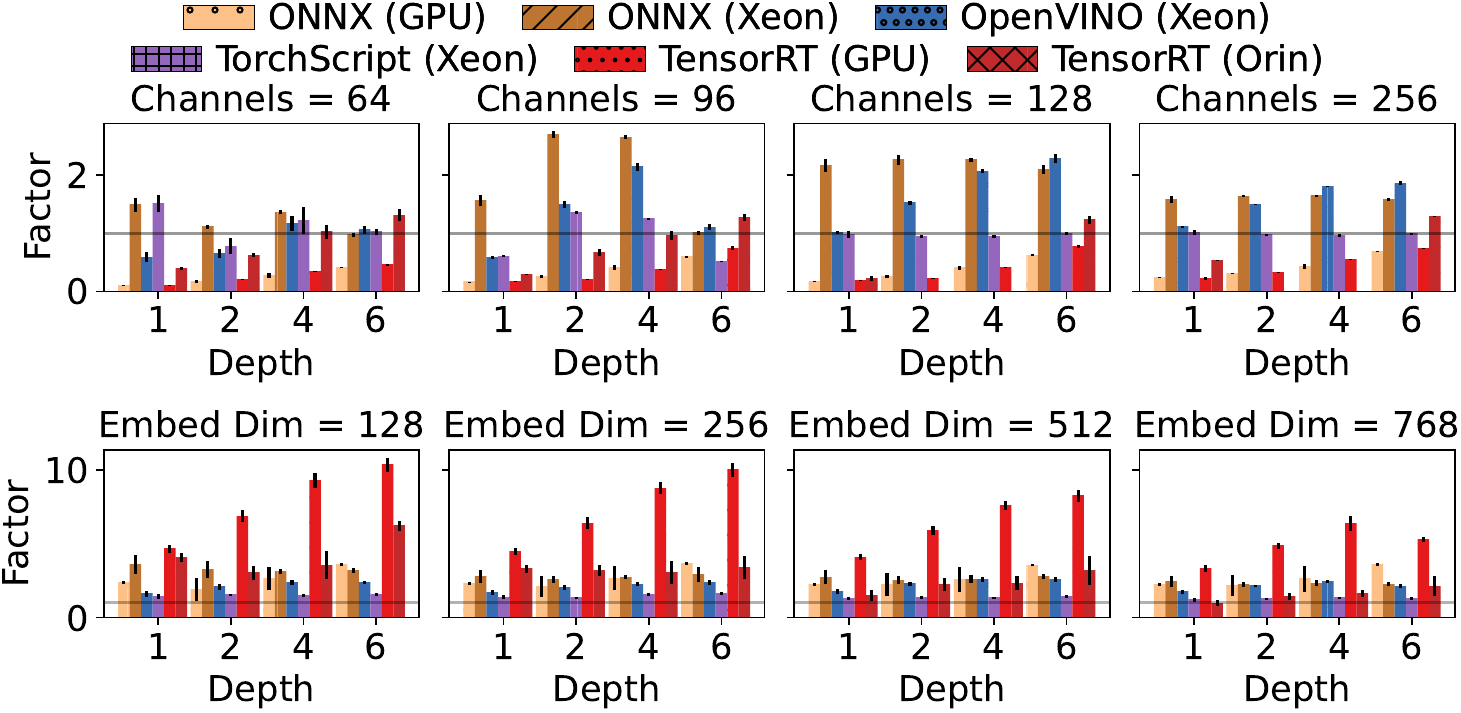}%
    }

    \vspace{1px} 

    \subfloat[Batch Size = 8\label{plot:speedup_throughput_inc_stacks_rel_bs8}]{%
        \includegraphics[width=\columnwidth]{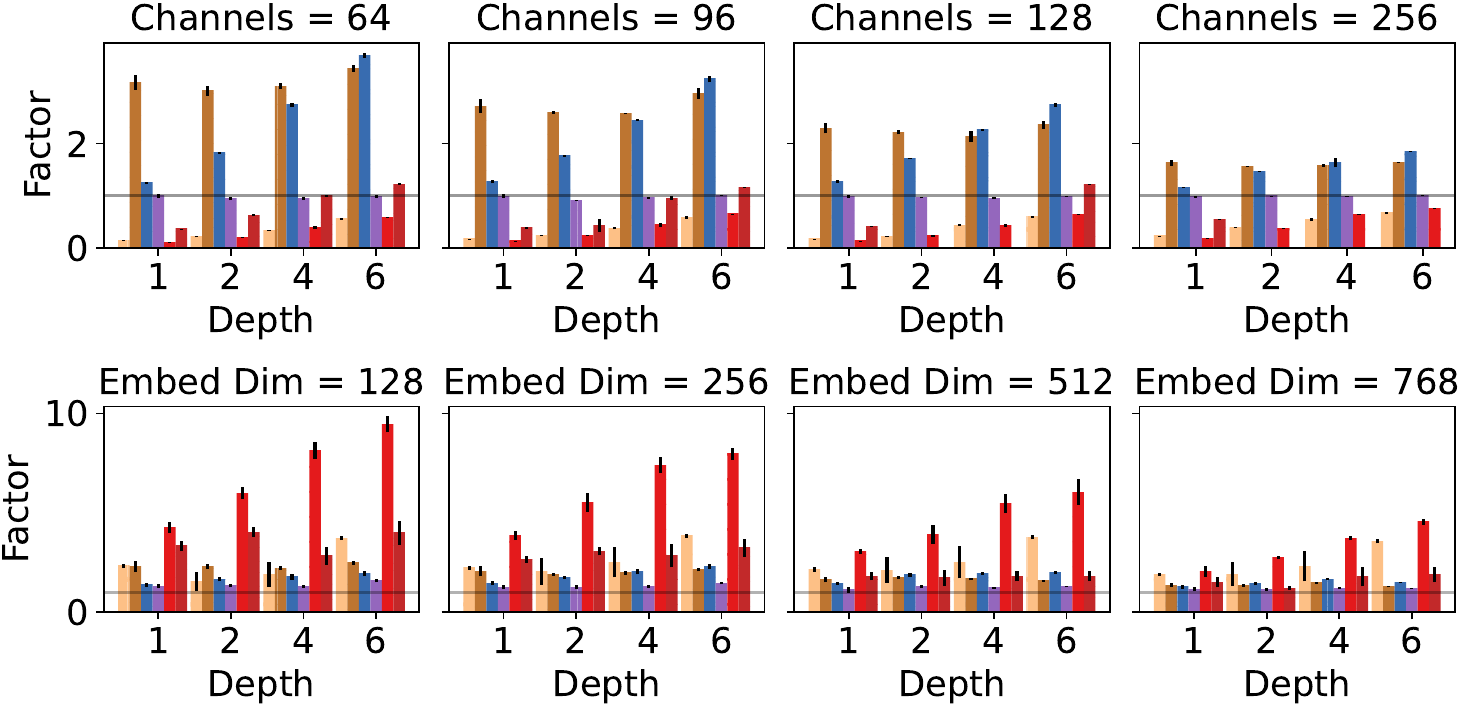}%
    }

    \vspace{1px} 

    \subfloat[Batch Size = 16\label{plot:speedup_throughput_inc_stacks_rel_bs16}]{%
        \includegraphics[width=\columnwidth]{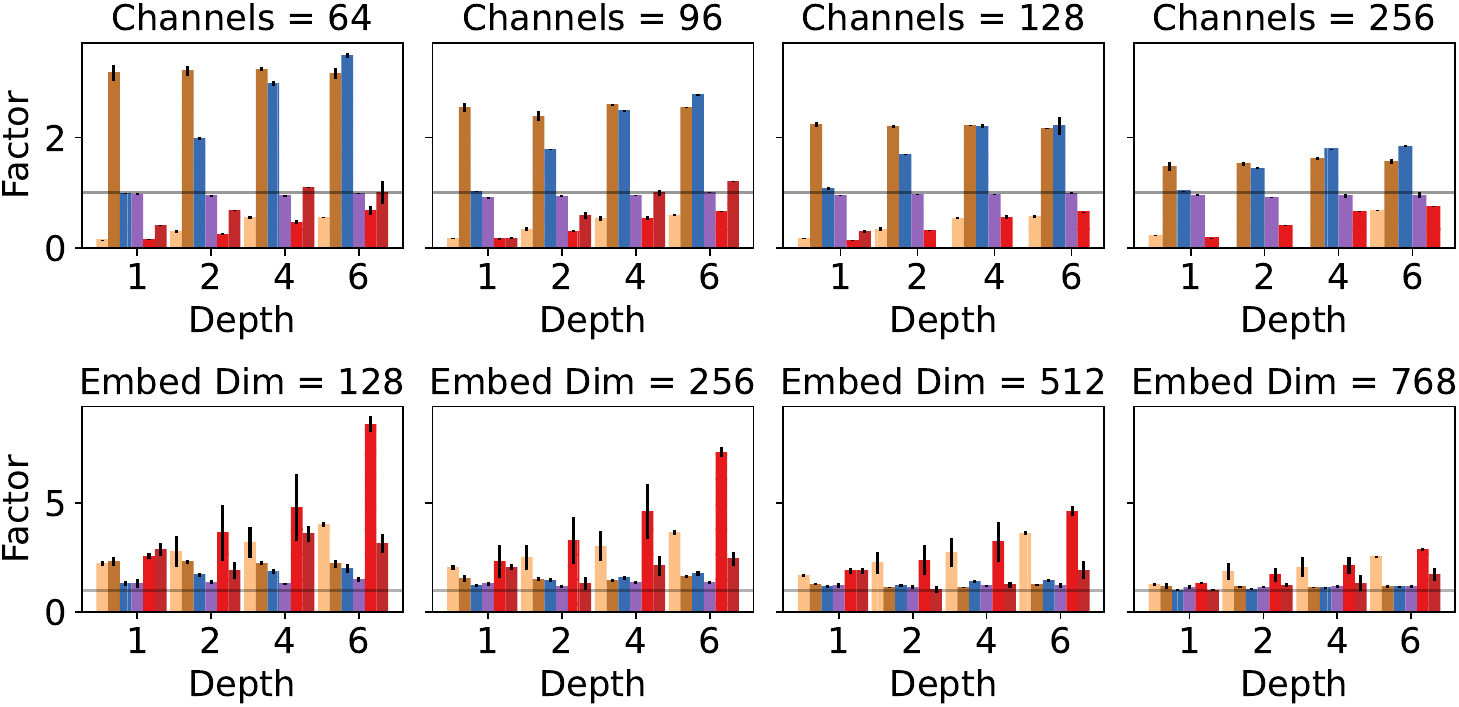}%
    }

    \caption{Contrasting how stacking homogeneous blocks improves relative throughput over the uncompiled graph.
    The relationship between depth and the factor remains consistent.
    } 
    \label{plot:speedup_throughput_inc_stacks_rel_combined} 
\end{figure}

\Cref{plot:speedup_throughput_inc_stacks_rel_combined} plots the throughput multiplier relative to the uncompiled graph for the convolutional and MHA blocks separately. Note that the Y-axis scaling is non-uniform to accentuate the relationship between compiler-device pairs at a set batch size. 

Unsurprisingly, we observe comparable behavior from convolutional and MHA blocks, similar to that of convolutional- and transformer-based architectures. However, the depth noticeably impacts the throughput relationship between the compilers. This trend persists even when performance begins to saturate due to the high computational load from large batch sizes and block widths. For example, the MHA block at a batch size of 8 with an embedding dimension of 128, using the TensorRT compiler on the GPU, achieves a throughput roughly twice that of the baseline dynamic computational graph on a single block, but increases to eightfold when stacking six repeating blocks. Conversely, convolutional architectures perform best using ONNX and OpenVINO on Xeon CPUs across the majority of our experiment configurations.
%
%
\Cref{tab:depth_scaling_conv} and \Cref{tab:depth_scaling_mha} quantify whether compilers can leverage the repeating patterns.
\begin{table}[htb]
\centering
\caption{Depth Scaling of Convolutional Blocks}
\label{tab:depth_scaling_conv}
\resizebox{\columnwidth}{!}{%
\begin{tabular}{lllrrrr}
Device & Compiler & Width & \multicolumn{2}{r}{Batch Size 8} & \multicolumn{2}{r}{Batch Size 16} \\ \cline{4-7} 
     &          &    & Slope   & Retention & Slope     & Retention \\ \hline
\rowcolor[HTML]{EFEFEF} 
GPU  & TensorRT & 64 & 0.08924 & 3.821     & 0.1032    & 1.187     \\
GPU  & TensorRT & 96 & 0.1024  & 0.2116    & 0.1038    & 0.6593    \\ \hline
\rowcolor[HTML]{EFEFEF} 
GPU  & ONNX     & 64 & 0.06953 & 1.153     & 0.09141   & 0.426     \\
GPU  & ONNX     & 96 & 0.07876 & 1.304     & 0.08367   & 0.7538    \\ \hline
\rowcolor[HTML]{EFEFEF} 
Orin & TensorRT & 64 & 0.1695  & 0.7846    & 0.1349    & 0.7759    \\
Orin & TensorRT & 96 & 0.1285  & 1.126     & 0.212     & 6.553     \\ \hline
\rowcolor[HTML]{EFEFEF} 
Xeon & ONNX     & 64 & 0.04862 & 1.009     & 0.0006561 & 0.9756    \\
Xeon & ONNX     & 96 & 0.0453  & 1.025     & 0.02658   & 0.9954    \\ \hline
\rowcolor[HTML]{EFEFEF} 
Xeon & OpenVINO & 64 & 0.4417  & 0.3379    & 0.4775    & 0.7428    \\
Xeon & OpenVINO & 96 & 0.3553  & 0.8669    & 0.3378    & 0.4106   
\end{tabular}%
}
\end{table}
\begin{table}[htb]
\centering
\caption{Depth Scaling of MHA Blocks}
\label{tab:depth_scaling_mha}
\resizebox{\columnwidth}{!}{%
\begin{tabular}{lllrrrr}
Device & Compiler & Width & \multicolumn{2}{r}{Batch Size 8} & \multicolumn{2}{r}{Batch Size 16} \\ \cline{4-7} 
       &          &       & Slope          & Retention       & Slope           & Retention       \\ \hline
GPU    & TensorRT & 128   & 0.8207         & 1.404           & 1.028           & 0.556           \\
\rowcolor[HTML]{EFEFEF} 
GPU    & TensorRT & 256   & 0.6369         & 0.6935          & 0.8831          & 1.471           \\ \hline
GPU    & ONNX     & 128   & 0.3594         & 2.187           & 0.3106          & 0.83            \\
\rowcolor[HTML]{EFEFEF} 
GPU    & ONNX     & 256   & 0.3406         & 1.724           & 0.2929          & 0.9108          \\ \hline
Orin   & TensorRT & 128   & 0.05291        & 1.178           & 0.173           & 0.9055          \\
\rowcolor[HTML]{EFEFEF} 
Orin   & TensorRT & 256   & 0.08368        & 0.95            & 0.1488          & 1.629           \\ \hline
Xeon   & ONNX     & 128   & 0.04244        & 1.105           & -0.01775        & 1.018           \\
\rowcolor[HTML]{EFEFEF} 
Xeon   & ONNX     & 256   & 0.01867        & 1.031           & 0.0215          & 0.9735          \\ \hline
Xeon   & OpenVINO & 128   & 0.1113         & 1.115           & 0.1355          & 0.9635          \\
\rowcolor[HTML]{EFEFEF} 
Xeon   & OpenVINO & 256   & 0.1589         & 1.577           & 0.1015          & 1.029          
\end{tabular}%
}
\end{table}

\noindent
We compute the \textit{slope} with the least‑squares fit to measure how much speed-up changes as we increase the depth. 

A positive slope implies that each additional block further increases the compiler's advantage. In contrast, a negative slope means that extra blocks diminish initial gains. The \textit{retention} is simply the ratio between the speed-up factor of the deepest stack and the speed-up factor of a single block.
A value close to 1 implies that a compiler cannot leverage the repeated block patterns, while values above 1 imply that the compiler can leverage repeated blocks. 
\Cref{tab:depth_scaling_conv} and \Cref{tab:depth_scaling_mha} should be interpreted jointly through slope and retention. The slope captures whether compiler speed-up increases as identical blocks are stacked, while retention captures how much of the single-block speed-up remains at the deepest stack. 
Together, they separate initial compilation gains from the extent to which those gains persist as depth increases. Using both values avoids treating a large single-block speed-up as evidence that compilation improves with depth. This distinction is important because shallow-block measurements can reflect fixed overhead reduction, while deeper stacks expose whether repeated composition changes the relative compiler advantage. These metrics are complementary. High initial speed-up can still yield modest retention, whereas lower initial speed-up can improve with depth. Across TensorRT configurations, convolutional blocks show higher average retention than MHA blocks, which is consistent with repeated Conv-BatchNorm-ReLU-like patterns being optimized more consistently on NVIDIA devices. On Xeon with OpenVINO, convolutional blocks also exhibit substantially larger positive slopes, indicating that deeper networks continue to expose optimization opportunities.
However, ONNX results and several MHA configurations retain values above 1, suggesting that their gains depend more on subgraph structure, tensor shapes, and attention-specific compiler support.

\bboxed{
Vendor-specific compilers can achieve increasing relative speed-up with depth, and we interpret this as evidence that repeated block structures create more favorable optimization opportunities for the evaluated compiler-device pairs. Repeated convolutional compositions show the most consistent gains, especially under TensorRT and OpenVINO. This is consistent with their regular Conv-BatchNorm-ReLU-like structure, which is commonly associated with effective fusion and scheduling. MHA blocks also benefit from compilation, but their gains are more configuration-dependent because the attention subgraph contains reshaping, matrix multiplication, softmax, residual, and normalization operations. The present measurements do not isolate the generated kernels, so our explanations should be read as interpretations of the observed behavior rather than as demonstrated compiler internals.
}
\subsection{Batch Parallelization Scaling Efficiency} \label{subsec:batchparanal}
\begin{figure}[htb]
    \centering
    \includegraphics[width=1\linewidth]{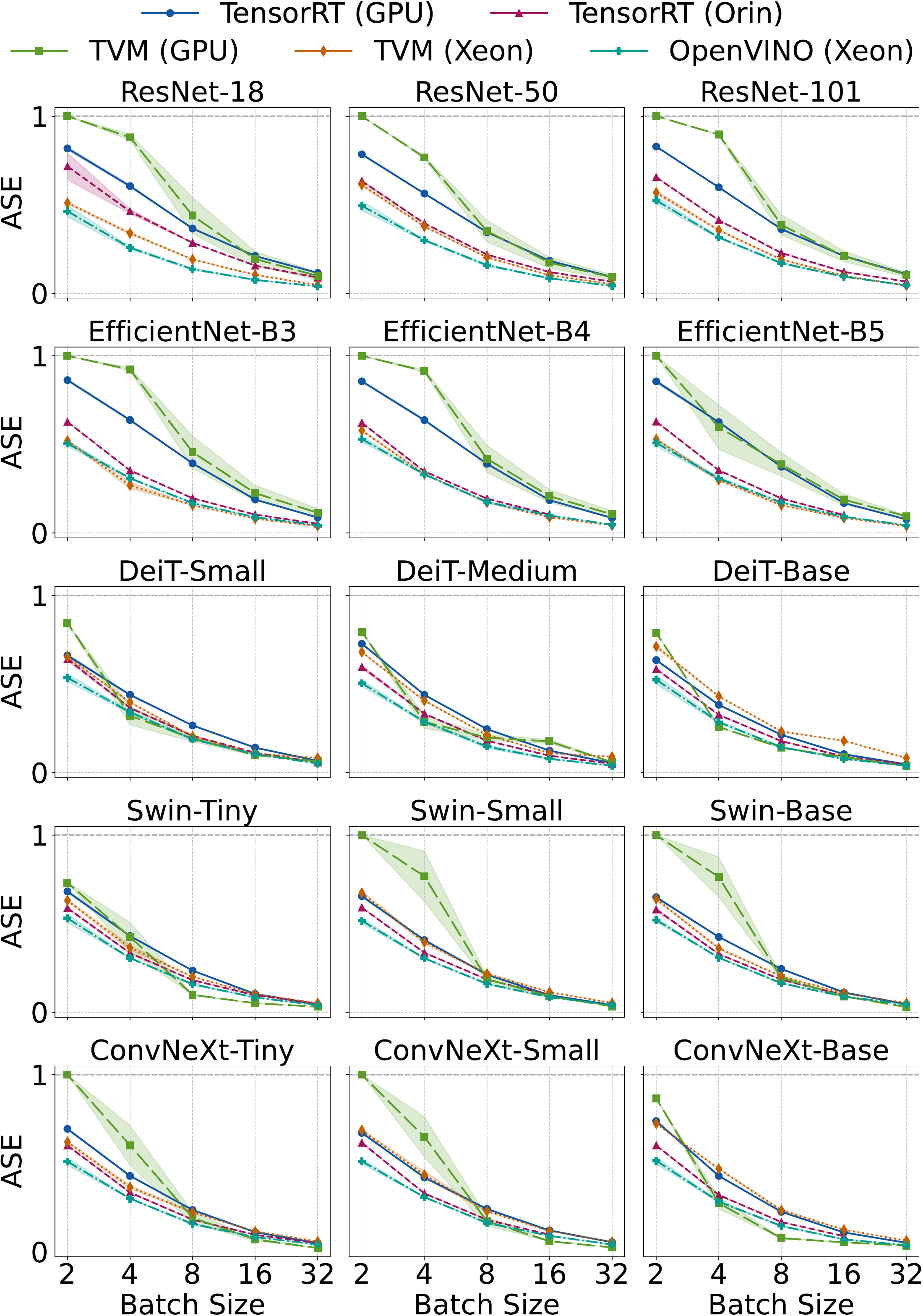}
    \caption{
    Absolute Scaling Efficiency (ASE) of architectural families. ASE is throughput at batch size b divided by b times throughput at batch size 1. Higher values indicate better batch scaling. ASE decreases as throughput saturates with batch size.
    }
    \label{plot:batchpar_network_ase}
\end{figure}

From both \Cref{plot:speedup_throughput_inc_network_rel} and
\Cref{plot:speedup_throughput_inc_stacks_rel_combined},
it is apparent that increasing the batch size significantly influences the throughput rate, and different compilers exhibit varying behavior.
\label{subsubsec:scalingeff}
\Cref{plot:batchpar_network_ase} illustrates explicitly how increasing the batch size decreases the scaling efficiency despite increasing the raw throughput. 

Apache TVM shows considerable but inconsistent scaling efficiency for convolutional-based architectures on the GPU. This is expected due to TVM's search-heuristic-based optimization, i.e., we must start a new search for each batch size. Conversely, the scaling efficiency decay of TensorRT is more predictable, as it is consistent with negligible variance. 
\begin{figure}[t]
    \centering
    \includegraphics[width=1\linewidth]{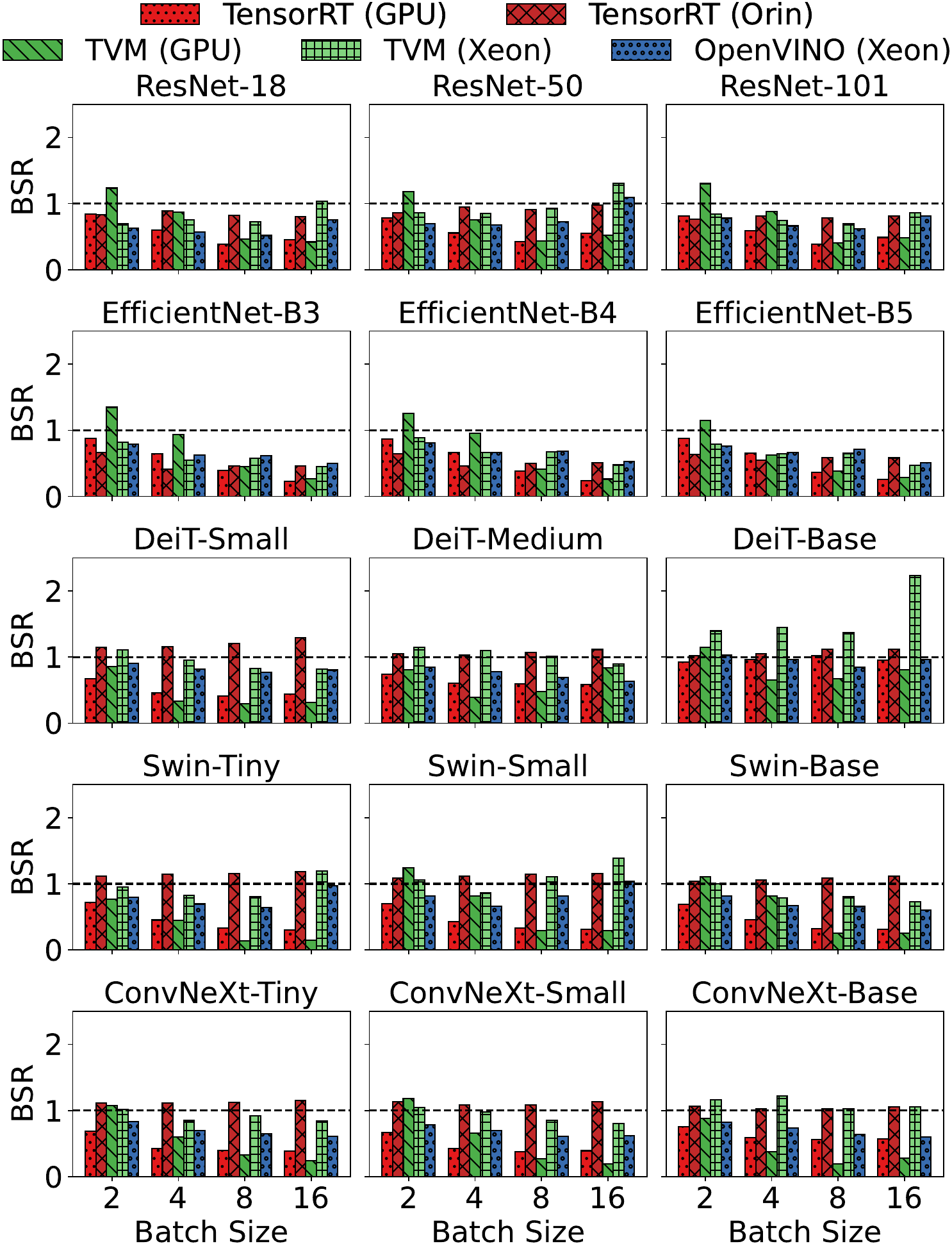}
    \caption{
    Batch Scaling Resilience (BSR) of architectural families. BSR is compiler ASE divided by dynamic-graph ASE. Values above 1 mean better batch-scaling efficiency than the baseline, while values below 1 mean steeper efficiency loss.
    }
    \label{plot:batchpar_bsr_networks}
\end{figure}
To account for varying compute capacities, and to provide information on relative improvement over the dynamic computational graph baseline, \Cref{plot:batchpar_bsr_networks} plots the batch scaling resilience (\Cref{subsubsec:measuringbatchpar}).
A BSR below 1.0 indicates that the compiled graph loses scaling efficiency faster than the dynamic graph under the same batch-size change. This, by itself, does not prove that the compiler missed a specific optimization. Rather, it identifies configurations where compilation improves absolute throughput but does not preserve batch-scaling efficiency relative to the baseline.
In particular, the erratic TVM results suggest that the fixed tuning budget does not consistently find schedules that preserve batch-scaling efficiency across model families and devices.
%
For example, applying TVM to the mid-sized DeiT model shows significantly higher resilience than OpenVINO on the Xeon CPU. Conversely, the resource-constrained Orin shows a BSR of roughly 1.0 across all transformer-based architectures and sizes, while achieving substantial throughput gains within the same architecture. However, especially for smaller architectures, the BSR of TensorRT on the powerful GPU is consistently below 1.0. 

\bboxed{The Batch Scaling Resilience (BSR) metric uncovers compiler-specific optimization patterns, demonstrating that TensorRT achieves consistent scaling profiles (BSR $\approx 1$) for most architectures while TVM shows erratic but occasionally superior resilience for certain configurations.
}
\subsection{Batch-Width Scaling Friction Mitigation} \label{subsec:frictionmitigation}
We investigate whether compilers can help address the batch-width scaling friction with block-level experiments for the same reasons as \Cref{subsec:depthscaling}.
\begin{figure}[htb]
    \centering 

    \subfloat[Convolutional Blocks (Depth=6) \label{plot:batchpar_ase_heatmaps_conv}]{%
        \includegraphics[width=\columnwidth]{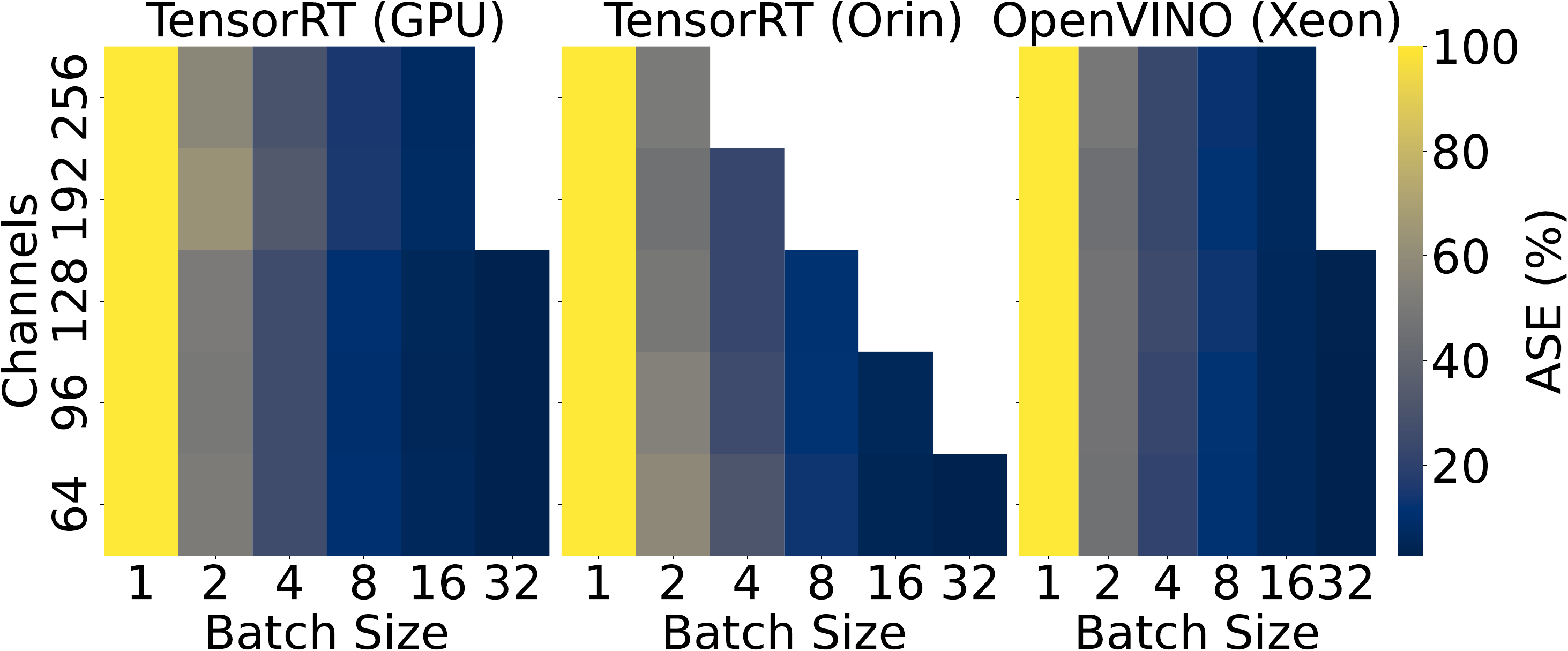}%
    } 

    \vspace{1ex} 

    \subfloat[MHA Blocks (Depth=6) \label{plot:batchpar_ase_heatmaps_mha}]{%
        \includegraphics[width=\columnwidth]{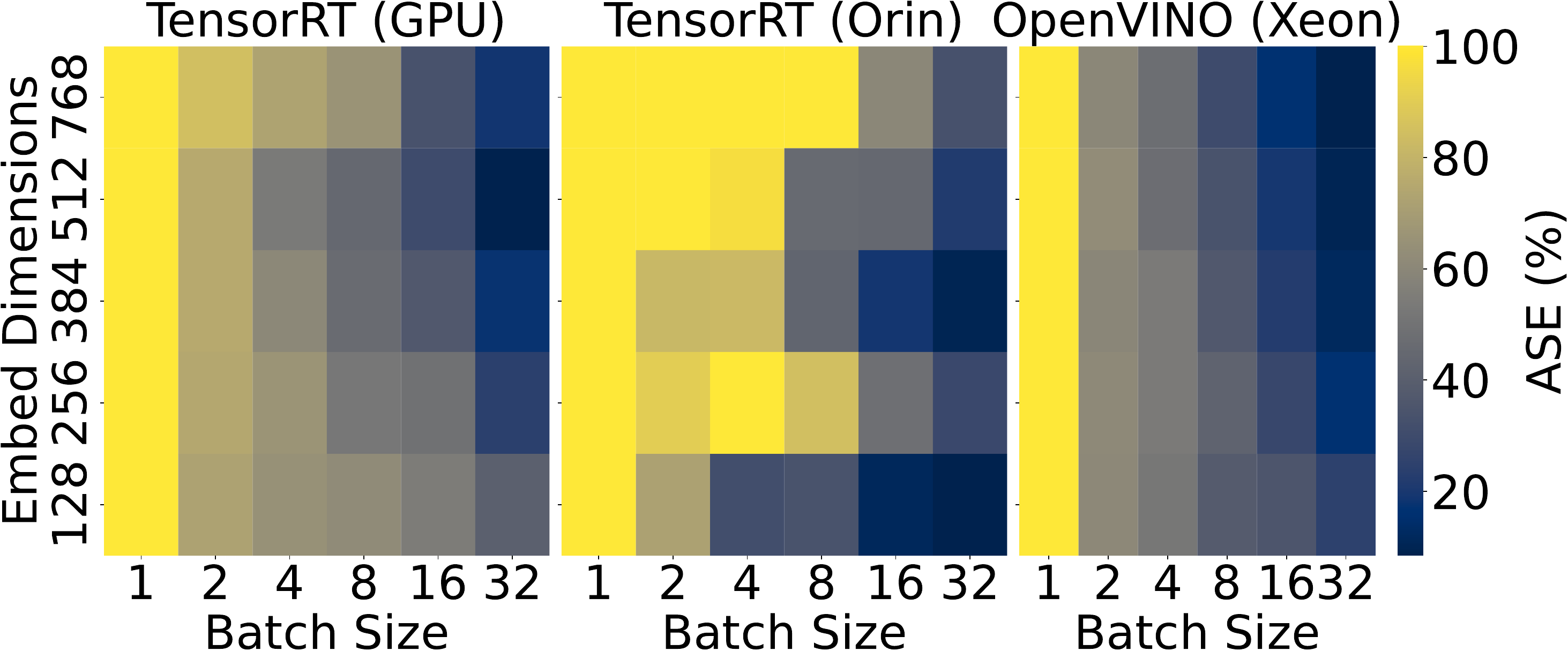}%
    }

    \caption{
    Effect of block width on Absolute Scaling Efficiency (ASE). Wider blocks reduce ASE more strongly, but the decrease depends on the compiler-device pair and block type.
    } 
    \label{plot:batchpar_ase_heatmaps_combined} 
\end{figure}

The batch-width friction is directly apparent from \Cref{plot:batchpar_ase_heatmaps_combined}.
As we increase the width of a block, the scaling efficiency drops considerably from one batch size to the next larger batch size.
\begin{figure}[htb]
    \centering 

    \subfloat[Depth = 1 \label{plot:batchpar_bsr_blocks_sz1}]{%
        \includegraphics[width=\columnwidth]{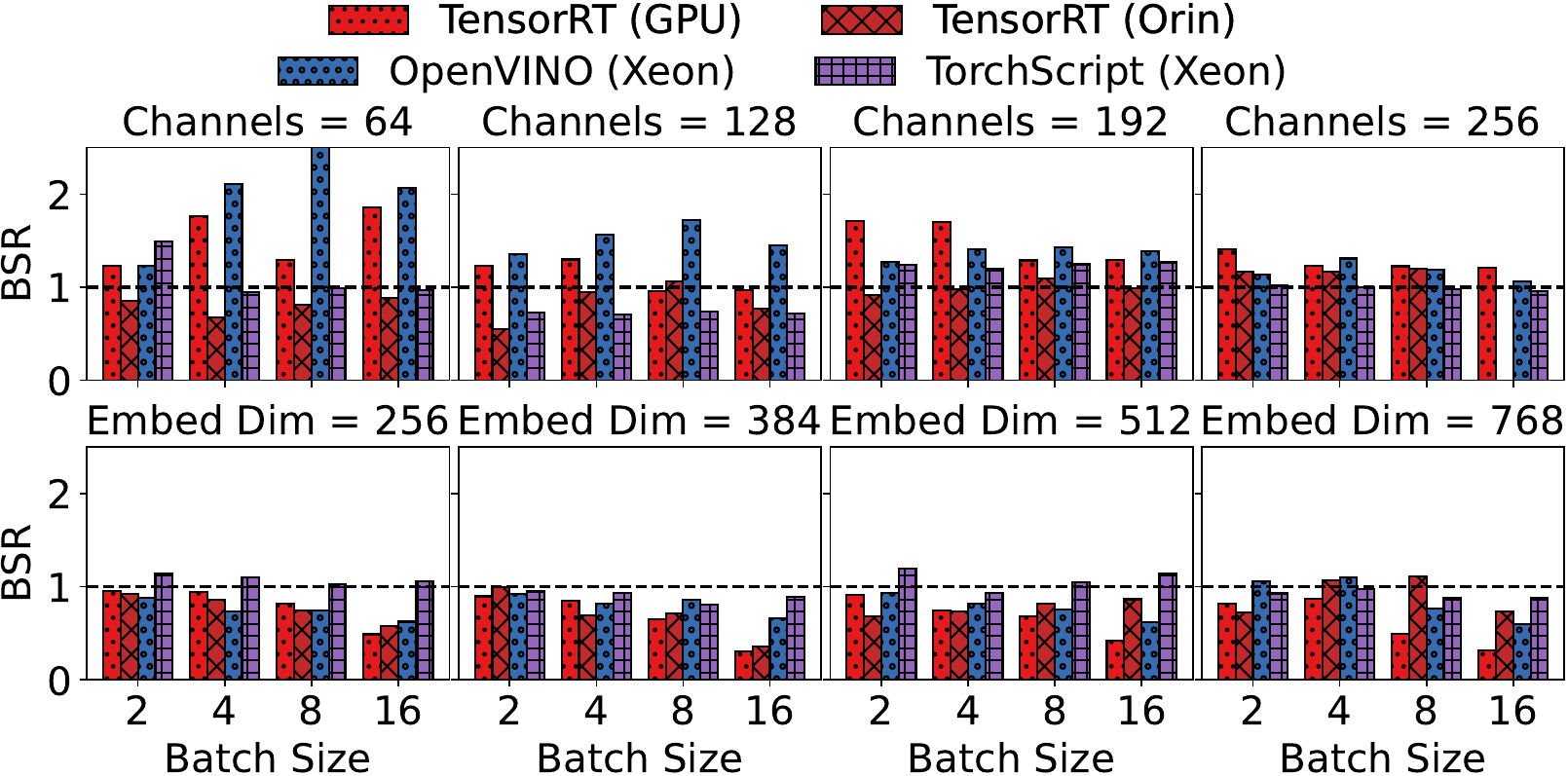}%
    } 


    \subfloat[Depth = 3 \label{plot:batchpar_bsr_blocks_sz2}]{%
        \includegraphics[width=\columnwidth]{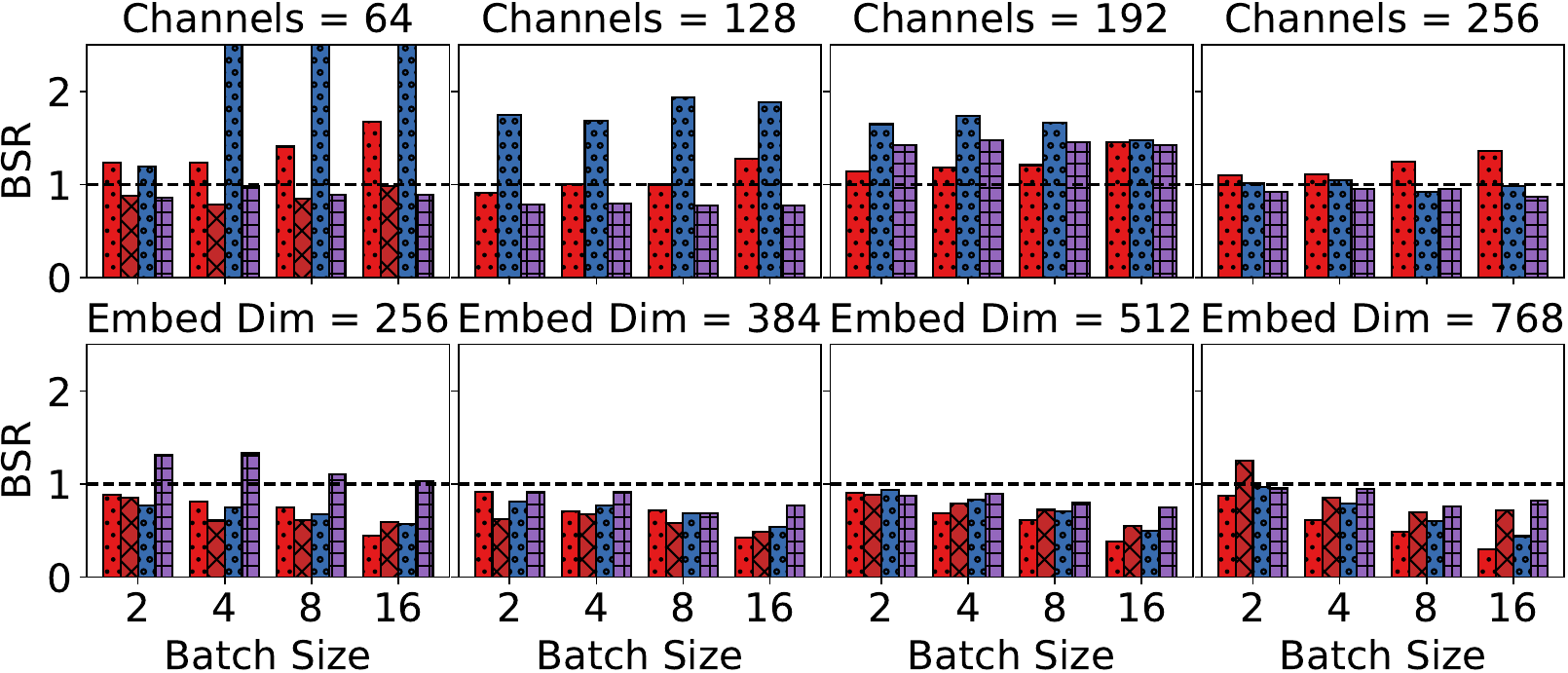}%
    } %


    \subfloat[Depth = 6 \label{plot:batchpar_bsr_blocks_sz3}]{%
        \includegraphics[width=\columnwidth]{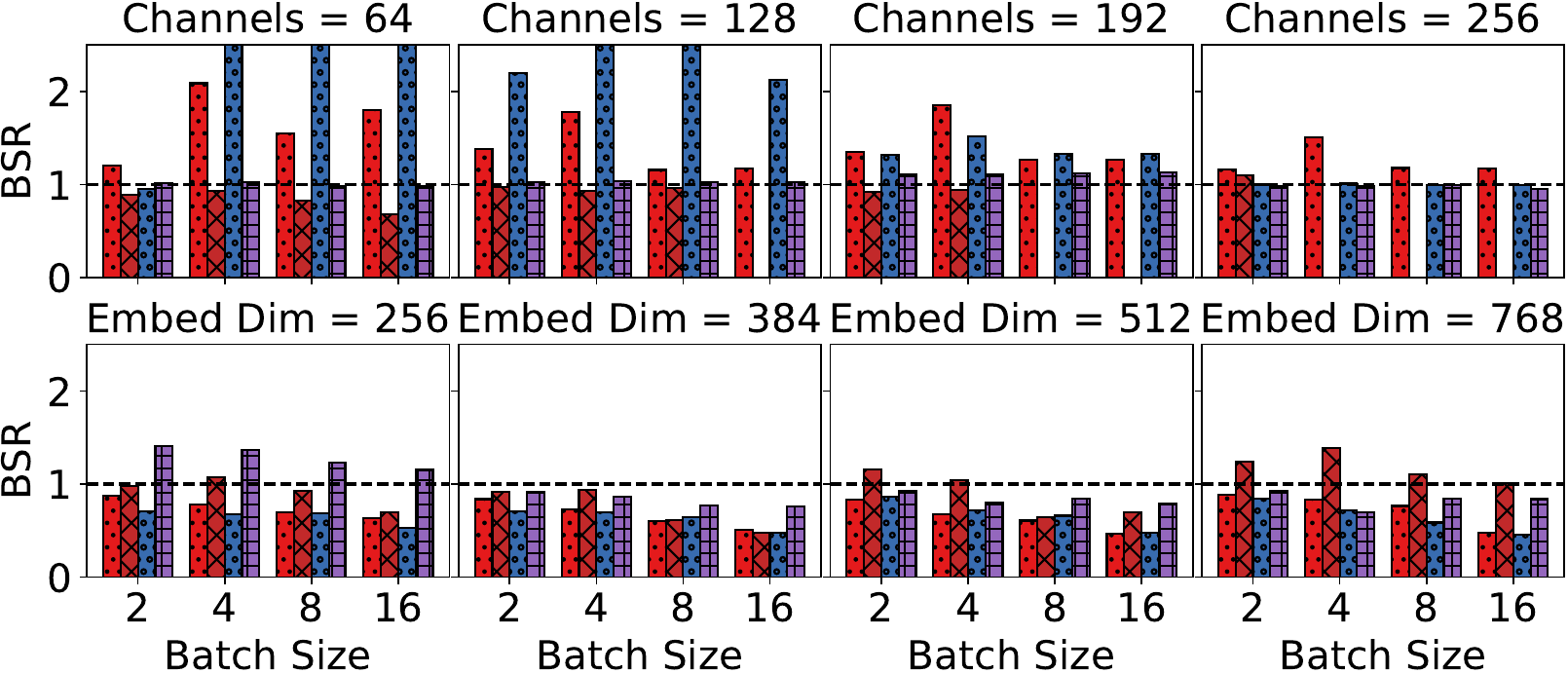}%
    } %
    \caption{
    Batch Scaling Resilience (BSR) for convolutional and MHA blocks. BSR compares compiled-graph ASE with dynamic-graph ASE. 
    Values above 1 indicate improved scaling resilience after compilation.
    } 
    \label{plot:batchpar_bsr_blocks} 
\end{figure}

\noindent
To compare with the baseline dynamic computational graphs and account for hardware differences, we run experiments across three configurations and plot the BSR in 
\Cref{plot:batchpar_bsr_blocks}.
Results show that increasing the depth tends to moderately improve BSR for TensorRT, likely for the same reasons outlined in \Cref{subsec:depthscaling}. TorchScript slightly mitigates the scaling friction for some configurations through intra-operator parallelization. 
A BSR value higher than one implies that scaling efficiency decreases more gracefully, while a BSR value lower than one indicates that compilation worsens the scaling efficiency degradation relative to the uncompiled dynamic graph.
%
This is best seen with OpenVINO. 
For the convolutional blocks, it can considerably mitigate the efficiency decrease, which is consistent with better use of the multiple CPU cores.

\bboxed{
Methods for resource-constrained environments should tune model width and batch size according to hardware characteristics when optimizing throughput. Compiler efficacy is crucial for mitigating batch-width scaling friction.
}
%
%
\subsection{Resource Usage Reduction} \label{subsec:loadreduction}
\noindent
From \Cref{tab:network_results_summary} it is apparent that when compilers successfully optimize the graph to have considerable throughput gains, the CPU usage increases on the Xeon where there is no dedicated GPU. On the GPU server and the Jetson board, TensorRT can decrease the CPU usage, marginally on the powerful server and significantly on the constrained Jetson board. 
TensorRT on GPU reduces CPU-side work during inference, consistent with static graph execution and reduced launch overhead. On the constrained Jetson Orin’s SoC, this appears as lower CPU usage.
On a higher-end GPU Server, these savings are negligible.
\begin{figure}[htb]
    \centering
    \includegraphics[width=\columnwidth]{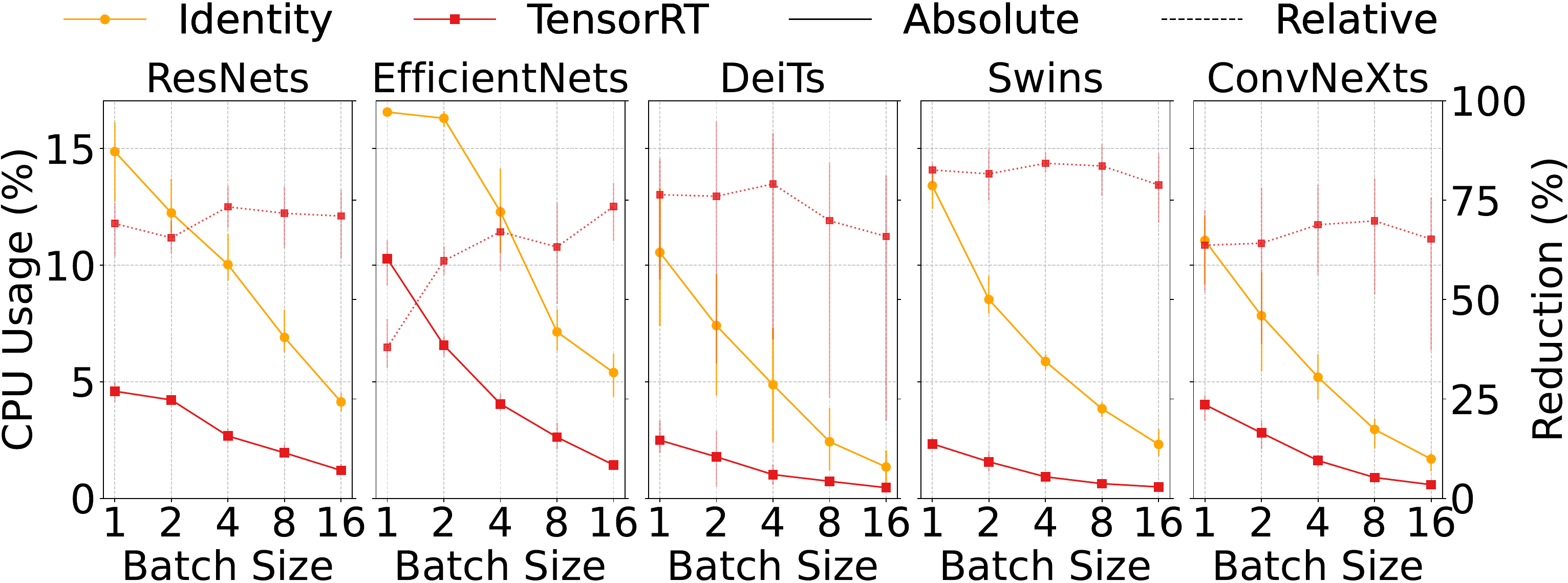}
    \caption{
    CPU usage and CPU-usage reduction on the Orin device. The left Y-axis shows absolute CPU usage for identity and TensorRT runs. The right Y-axis shows relative CPU-usage reduction from TensorRT. As batch size increases, throughput saturates on the GPU and CPU usage decreases.
    }
    \label{plot:resource_usage_orin_cpu_abs}
\end{figure}
\Cref{plot:resource_usage_orin_cpu_abs} directly compares the CPU usage of TensorRT on the compiled network and the baseline on all architectural families on the Orin device. Notice that, for transformer-based architectures, particularly the Swin family, CPU usage can be reduced by up to 80\%. This is particularly valuable in constrained environments that require auxiliary tasks to be performed on the CPU. For example, in \cite{fool}, interference from pre- and post-processing on the CPU was negligible on smaller models but adversely affected the throughput of larger models.
%

\bboxed{
Applying compilers on devices with a dedicated accelerator can significantly reduce CPU utilization consistently in our measurements. These findings indicate compiler selection should consider both throughput and resource utilization metrics.
CPU usage reduction is particularly valuable in edge computing scenarios where horizontal scaling is limited, and CPUs may handle concurrent auxiliary tasks. 
}
\subsection{Discussion }\label{subsec:discussion}
A limitation of this evaluation is that compiler versions and compiler configurations are experimental factors. The measurements in this paper reflect a representative default setup. They should therefore not be read as definitive statements about TensorRT, OpenVINO, TVM, ONNX, or TorchScript in general. Different compiler releases, tuning budgets, precision settings, calibration procedures, or target-specific flags may change the absolute throughput and some relative rankings. The intended use of the results is to show how such changes can affect ML systems' conclusions and why compiler-aware benchmarking should be part of the research workflow.
Our findings are largely consistent with prior benchmarking work, though our methodological scope differs. Earlier studies typically focus on isolated compilers or per-operation performance, whereas our evaluation integrates compiler effects into end-to-end model benchmarking. Consistent with Zhou \& Yang \cite{tensorrteval} and Li et al. \cite{dlcompilersurvey}, we observe that vendor-specific compilers achieve superior throughput due to aggressive fusion of supported operator patterns. However, unlike Xing et al. \cite{comparisononhardware}, who report stable performance ordering across architectures, our cross-compiler evaluation shows that compilation can invert relative throughput rankings when optimization coverage diverges. This apparent inconsistency arises from differences in experimental design, explained by our methodology using fully compiled workloads on heterogeneous edge--cloud hardware rather than operation-level microbenchmarks. Moreover, while Zhang et al. \cite{mobilelib, mobilelib2} attribute performance variance across libraries to framework differences, our results subsume their finding that underlying compiler optimizations are the decisive factor. In summary, our work complements earlier studies by extending the benchmarking dimension from the framework level to the compiler level and providing diagnostic metrics, such as BSR, for future analyses.
\section{Conclusion} \label{sec:conclusion}
We presented a methodology that complements ongoing compiler advances by providing a measurement-driven baseline for ML researchers while compiler support continues to evolve.
The contribution is methodological and empirical: it quantifies how compilation can overturn pre-compilation performance expectations and provides metrics to diagnose scaling behavior. The BSR metric quantifies a compiler's ability to mitigate performance friction as batch size increases. Block-level experimentation confirmed that simple compositions with widely supported operations provide significant advantages in resource-constrained environments, as compilers effectively leverage repeated patterns for disproportionate throughput gains. We emphasize that the evaluation relies on a finite heterogeneous testbed, which does not capture the full diversity of emerging accelerator architectures. Scaling the methodology to distributed settings or deployments that include model compression is left for future investigation.
%
\section*{Acknowledgment}
We thank Alexander Knoll for providing us with the hardware infrastructure.
\bibliographystyle{ieeetr}
\bibliography{main}
\vspace{-15mm}
\begin{IEEEbiography}[{\includegraphics[width=1in,height=1.25in,clip,keepaspectratio]{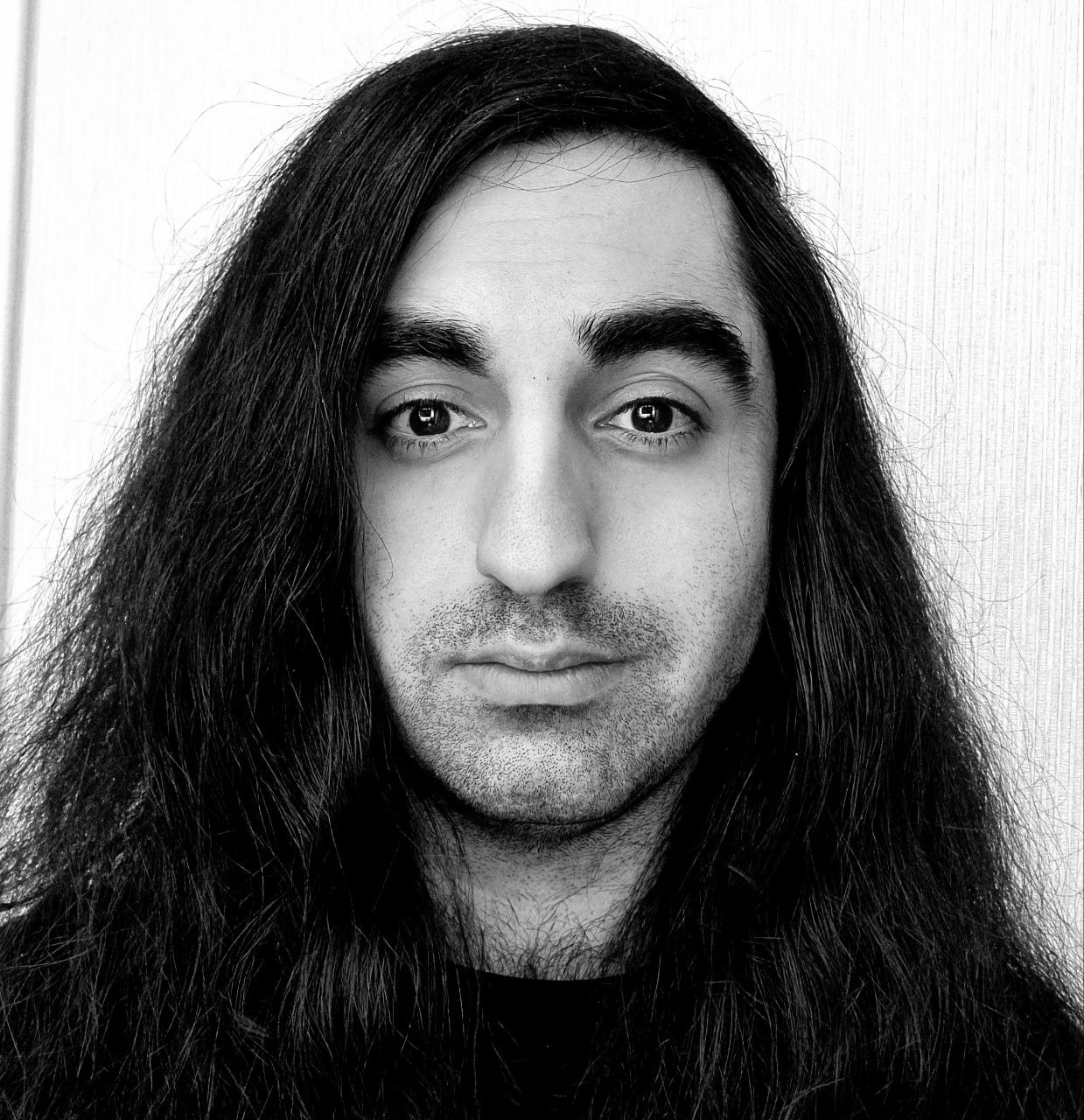}}]{Alireza Furutanpey} received a MSc and PhD from the Technical University of Vienna, Austria, in 2022 and 2025, respectively, with distinction in the field of Computer Science. He is currently leading the research efforts at Coovally in Barcelona and is an external scientist at the Distributed Systems Group in Vienna, focusing on Edge Computing. His research interests include Mobile Edge Computing, Edge Intelligence, and Machine Learning.
\end{IEEEbiography}
\vspace{-15mm}
\begin{IEEEbiography}[{\includegraphics[width=1in,height=1.25in,clip,keepaspectratio]{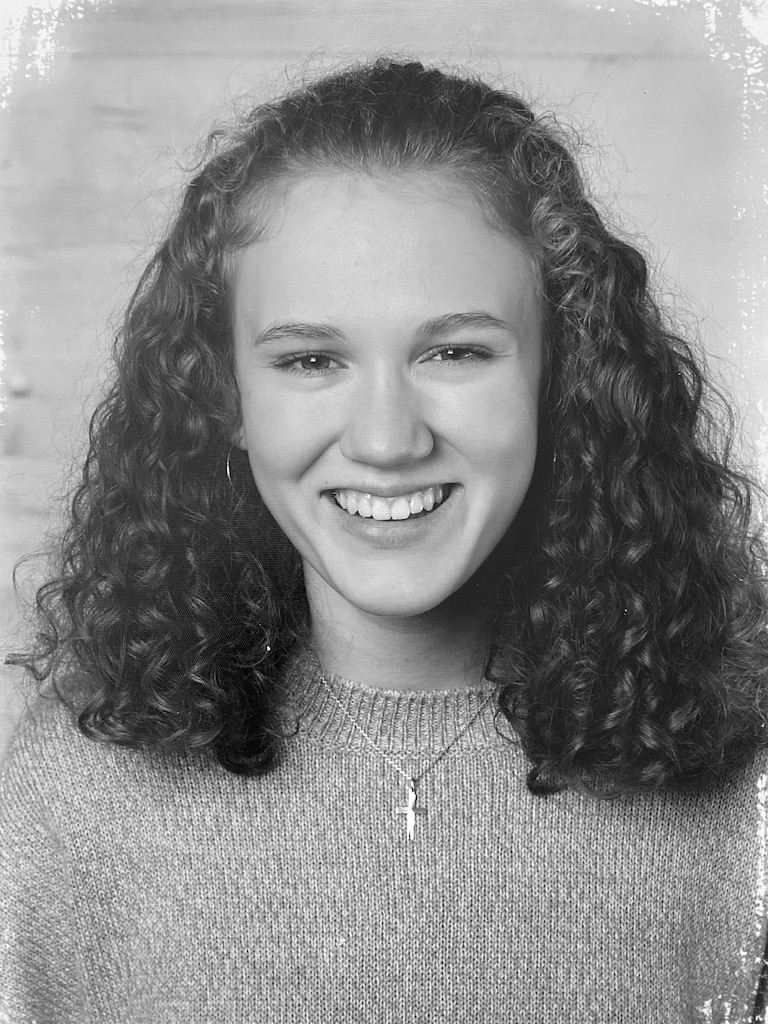}}]{Carmen Walser} is a Master's student at the Technical University of Vienna. In 2024, she received the Siemens Awards for Excellence. Her research interests include Edge Computing, Low-level optimization, and IoT systems. 
\end{IEEEbiography}
\vspace{-15mm}
\begin{IEEEbiography}[{\includegraphics[width=1in,height=1.25in,clip,keepaspectratio]{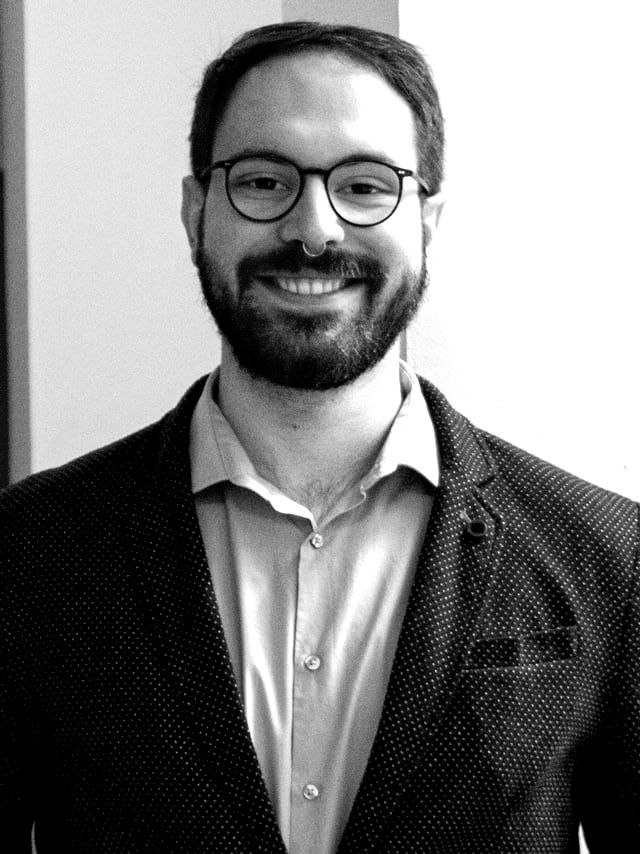}}]{Philipp Raith} received a MSc and PhD from the Technical University of Vienna, Austria, in 2021 and 2025, respectively, with distinction in the field of Computer Science.  He is currently leading engineering efforts at Coovally in Barcelona and is an external scientist at the Distributed Systems Group in Vienna. His research interests include Serverless Edge Computing, Edge Intelligence, and Operations for AI.
\end{IEEEbiography}
\vspace{-15mm}
\begin{IEEEbiography}[{\includegraphics[width=1in,height=1.25in,clip,keepaspectratio]{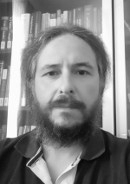}}]{Pantelis A. Frangoudis} is a researcher with the Distributed Systems Group, TU Wien, Austria. He has been with the Communication Systems Department, EURECOM, France (2017–2019), and with team DIONYSOS at IRISA/INRIA Rennes, France (2012–2017)., which he originally joined under an ERCIM ``Alain Bensoussan'' post-doctoral fellowship. He has a Ph.D. (2012) in Computer Science from AUEB, Greece. His interests include mobile and wireless networking, edge and cloud computing, and Internet multimedia.
\end{IEEEbiography}
\vspace{-15mm}
\begin{IEEEbiography}[{\includegraphics[width=1in,height=1.25in,clip,keepaspectratio]{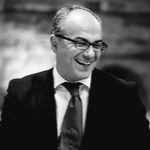}}]{Schahram Dustdar} is a full professor of computer science and heads TU Wien's Distributed Systems Group. His research interests include distributed systems, Edge Intelligence, complex and autonomic software systems. He's the editor-in-chief of Computing;  associate editor of ACM Transactions on the Web, ACM Transactions on Internet Technology, IEEE Transactions on Cloud Computing, and IEEE Transactions on Services Computing. He's also on the editorial boards of IEEE Internet Computing and IEEE Computer. He has received the ACM Distinguished Scientist award, the Distinguished Speaker Award, and the IBM Faculty Award. He is an elected member of Academia Europaea, where he was Informatics Section chairman from 2015 to 2022. He is an IEEE Fellow and AAIA Fellow where he is the current President.
\end{IEEEbiography}
\end{document}